\documentclass[11pt]{article}

\usepackage{arxiv}

\usepackage[utf8]{inputenc} 
\usepackage[T1]{fontenc}    
\usepackage[colorlinks=true, linkcolor=blue, citecolor=blue, urlcolor=blue, breaklinks=true, pdfborder={0 0 0}]{hyperref}       
\usepackage{url}            
\usepackage{booktabs}       
\usepackage{amsfonts}       
\usepackage{nicefrac}       
\usepackage{microtype}      
\usepackage{graphicx}
\usepackage{doi}

\usepackage{amsmath}
\usepackage{array}

\usepackage{hyperref}
\usepackage{natbib}
\usepackage[table]{xcolor}
\definecolor{highlight}{HTML}{D1D1D1}

\AtBeginEnvironment{appendix}{
  \counterwithin{figure}{section}
}

\title{bayesNMF: Fast Bayesian Poisson NMF with Automatically Learned Rank Applied to Mutational Signatures}

\author{ 
    \href{https://orcid.org/0000-0002-1411-8750}{\includegraphics[scale=0.06]{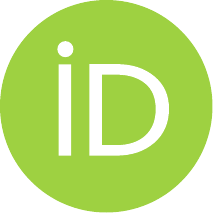}\hspace{1mm}\textcolor{black}{Jenna M. Landy}}\\
	Harvard University \\
	\texttt{jlandy@g.harvard.edu} \\
    \And
    {Nishanth Basava}\\
	 Vanderbilt University\\
	\texttt{nishanth.r.basava@vanderbilt.edu} \\
    \And
    \href{https://orcid.org/0000-0002-8783-5961}{\includegraphics[scale=0.06]{orcid.pdf}\hspace{1mm}\textcolor{black}{Giovanni Parmigiani}} \\
	Dana Farber Cancer Institute\\
        Harvard University\\
        \texttt{gp@jimmy.harvard.edu}
}

\renewcommand{\shorttitle}{bayesNMF: Fast Bayesian Poisson NMF}

\usepackage{setspace}
\usepackage{caption}
\usepackage{setspace}
\usepackage{cleveref}
\crefname{appendix}{}{}

\usepackage{chngcntr} 

\begin{document}
\date{}
\maketitle
\begin{abstract}
Bayesian Poisson Non-Negative Matrix Factorization (NMF) is widely used to model count data, including in cancer mutational signature analysis. However, standard Gibbs samplers rely on computationally expensive Poisson augmentation, and current software implementations learn the latent rank either through slow and potentially subjective heuristic rank selection or with automatic approaches that do not report posterior uncertainty. In this paper, we introduce bayesNMF, an MH-within-Gibbs sampler to address both of these limitations. First, we define high-overlap proposals for Metropolis-Hastings sampling to remove the need for Poisson augmentation. Second, we define a BIC-based sparsity prior to learn rank automatically within the Bayesian formulation while allowing for posterior uncertainty quantification. We provide an open-source R software package with all of the models and plotting capabilities demonstrated in this paper on GitHub at \href{https://github.com/jennalandy/bayesNMF}{jennalandy/bayesNMF}. Although our applications focus on cancer mutational signatures, our software and results can be extended to any use of Bayesian Poisson NMF.
\end{abstract}


\raggedbottom

\section{Introduction}
\label{sec:intro}

Non-Negative Matrix Factorization (NMF) is a widely used unsupervised technique for dimension reduction and intuitive parts-based representation \citep{lee1999learning}, and is typically formulated with either a Poisson or a Normal likelihood. In cancer genomics, Poisson NMF applied to mutation counts assumes that tumor genomes arise from multiple additive mutational processes, with the resulting latent factors termed mutational signatures \citep{alexandrov2013signatures}. Within this framework, Bayesian NMF allows for integration of prior knowledge and posterior uncertainty quantification \citep{cemgil2009, schmidt2009}.

The computational cost of a Gibbs sampler for Bayesian NMF grows with the number of samples $G$ (columns), latent factors $N$, and variables $K$ (rows). A Normal likelihood requires approximately $N(K + G)$ updates per iteration \citep{schmidt2009}. A Poisson likelihood, however, relies on Poisson augmentation for conditional updates to follow standard distributions. Each observed count is decomposed into latent factor-specific counts, introducing $NKG$ auxiliary parameters. This requires $N(K+G+KG)$ updates per iteration \citep{cemgil2009}, resulting in a substantially higher computational cost. With growing publicly available data sources \citep{international2010international, weinstein2013cancer, icgc2020pan} and an expanding set of previously discovered signatures \citep{alexandrov2020repertoire}, computational efficiency is a central constraint in large-scale mutational signature analysis.

Further, identifying latent rank is a fundamental aspect of any factor model. Most Bayesian NMF methods take a heuristic approach, fitting a separate model for each rank and optimizing regularized metrics such as the Bayesian information criterion (BIC) \citep{rosales2016, gori2018sigfit} or determining the ``elbow'' of metrics such as MSE or cosine similarity \citep{gori2018sigfit, islam2022uncovering}. This process is computationally intensive and relies on post hoc and potentially subjective model selection.

Existing methods address subsets of these challenges, but trade off posterior uncertainty, computational efficiency, and rank learning in different ways (Table \ref{tab:software}). SignatureAnalyzer \citep{kasar2015whole} uses automatic relevance determination (ARD) to learn rank as part of the Bayesian model. It avoids Poisson augmentation by optimizing the posterior numerically, but as a result, it provides only point estimates without posterior uncertainties. On the other hand, SigFit \citep{gori2018sigfit} avoids Poisson augmentation by performing posterior inference in Stan \citep{carpenter2017stan}, which implements Hamiltonian Monte Carlo \citep{neal2011mcmc}, but selects rank heuristically across candidate models. Compressive NMF \citep{zito2024compressive} learns rank automatically while maintaining posterior uncertainty, but relies on computationally expensive Poisson augmentation. 

To our knowledge, no method simultaneously 1) avoids Poisson augmentation, 2) learns rank within the Bayesian formulation, and 3) reports posterior uncertainty. This paper introduces two innovations to address this gap. First, we define high-overlap proposals for Metropolis-Hastings-within-Gibbs sampling to avoid Poisson augmentation. Second, we define a BIC-based sparsity prior to learn rank within the Bayesian model. The accuracy and efficiency of the combined approach are evaluated through simulation studies, and results are presented on 32 cancer types from the Pan-Cancer Analysis of Whole Genomes (PCAWG) database \citep{icgc2020pan}. Methods are implemented in an open-source R package, {\tt bayesNMF}, available on GitHub at \href{https://github.com/jennalandy/bayesNMF}{jennalandy/bayesNMF}. 

\section{Background}\label{sec:background} 

\begin{table}
    \centering
    \begin{tabular}{|c||c|c|c|c|c|}
        \hline
        \textbf{Software} & \textbf{Automatically} & \textbf{Posterior} & \textbf{Avoids Poisson}\\
        &\textbf{Learns Rank}&\textbf{Uncertainty}&\textbf{Augmentation}\\
        \hline
        \rowcolor{highlight}{\tt bayesNMF} (this work) & $\checkmark$ & $\checkmark$ & $\checkmark$ \\

        \hline
         {\tt SignatureAnalyzer} (2015) & $\checkmark$ &  & $\checkmark$ \\
         \hline
         {\tt SigneR} (2016)& & $\checkmark$ & \\
         \hline
         {\tt SigFit} (2018) & & $\checkmark$ & $\checkmark$ \\
         \hline
         {\tt CompNMF} (2024)&$\checkmark$&$\checkmark$&\\
         \hline
    \end{tabular}
    \caption{Software for Bayesian Poisson NMF applied to de-novo mutational signature analysis: {\tt SignatureAnalyzer} \citep{kasar2015whole, kim2016somatic}, {\tt SigneR} \citep{rosales2016}, {\tt SigFit} \citep{gori2018sigfit}, and {\tt CompNMF} \citep{zito2024compressive}. An ideal software would automatically learn rank within the Bayesian formulation, report posterior uncertainty, and avoid computationally intensive Poisson augmentation.}
    \label{tab:software}
\end{table}

\subsection{Non-Negative Matrix Factorization (NMF) and Bayesian NMF} \label{sec:background-nmf}

NMF is an unsupervised method to decompose non-negative data $M$ ($K$ variables as rows by $G$ samples as columns) into two lower-rank non-negative matrices, $P$ and $E$ \citep{paatero1994positive, lee1999learning}. NMF represents each observation of the data matrix (column of $M$) as a linear combination of latent factors (columns of $P$):
\begin{align*}
    M_{kg} &= \sum_{n = 1}^N P_{kn}E_{ng}, \hspace{4mm} 
    \mbox{i.e.,} \;\; M = PE,\hspace{4mm} M \in \mathbb{R}_{\ge 0}^{K\times G}, P \in \mathbb{R}_{\ge 0}^{K\times N}, E \in \mathbb{R}_{\ge 0}^{N\times G}.
\end{align*}

NMF is typically fit by minimizing a reconstruction error, often the generalized Kullback-Leibler (KL) divergence (Equation~\ref{eq:kl}) or the Frobenius norm (Equation~\ref{eq:frob}). 
\begin{align}
    KL(M||PE) &= \sum_{kg} \left(M_{kg}\log\frac{M_{kg}}{(PE)_{kg}} - M_{kg} + (PE)_{kg}\right)\label{eq:kl}\\
    ||M-PE||_F^2 &= \sum_{kg}(M_{kg} - (PE)_{kg})^2 \label{eq:frob}
\end{align}
For each, \citet{lee2000algorithms} adapt gradient descent by deriving a step size that preserves nonnegativity, resulting in multiplicative update algorithms. These equate to probabilistic models: minimizing KL-divergence is equivalent to maximizing a Poisson likelihood on $M$ with mean $PE$ \citep{cemgil2009}, and minimizing the Frobenius norm is equivalent to maximizing a Normal likelihood on $M$ with mean $PE$ \citep{schmidt2009}.

Bayesian approaches place priors on $P$ and $E$ \citep{cemgil2009, schmidt2009}. Sampling the posterior distribution of Bayesian Poisson NMF with a standard Gibbs sampler requires Poisson augmentation \citep{cemgil2009}. This introduces latent variables $Z_{kng}$, defined as the counts from $M_{kg}$ attributed to latent factor $n$. Each $Z_{kng}$~follows a Poisson distribution with mean $P_{kn}E_{ng}$ and satisfies $M_{kg} = \sum_n Z_{kng}$. These additional $K N G$ parameters increase the computational requirements of Bayesian Poisson NMF far beyond those of Bayesian Normal~NMF.

\subsection{Cancer Mutational Signatures}\label{sec:background-mutsig}

Cancer is driven by genetic changes, including somatic DNA mutations, which can be caused by multiple processes. Understanding which mutational process activities are reflected in a tumor genome can help answer questions about cancer development, subtyping, prognosis, and treatment \citep{nikzainal2012_mutproc, nik2012life, alexandrov2013signatures, rosales2016}. A computationally-derived mutational signature models the specific patterns produced by a mutational process. Signatures have previously been linked to DNA damage repair deficiencies \citep{zamborszky2017loss}, deamination of 5-methylcytosine \citep{nikzainal2012_mutproc}, tobacco smoke \citep{alexandrov2013signatures, nik2015genome}, and ultraviolet radiation \citep{nik2015genome, saini2016impact, hayward2017whole}. The Catalog of Somatic Mutations in Cancer (COSMIC) \citep{tate2019cosmic} provides a high-confidence set of reference signatures. These are an important reference, but cannot be treated as ground truth, as they were estimated from data.

Mutational signatures are learned with NMF or Bayesian NMF on a count matrix of~$K$ mutation types by $G$ samples or tumor genomes. This results in a signatures matrix, $P$, holding (potentially unnormalized) relative frequencies with which each signature gives rise to each mutation type, and an exposures matrix, $E$, with relative contributions of each signature to each sample. Single base substitution (SBS) mutations are often categorized by the substitution and its immediate left and right nucleotide bases (e.g., A[C$>$T]G). Base pairing maps mutations to originate with either Thymine (T$>$$\cdot$) or Cytosine (C$>$$\cdot$), resulting in 96 mutation types. The standard model for mutational signature analysis is a Poisson likelihood. Most Bayesian models use Gamma priors \citep{rosales2016, grabski2025bayesian, zito2024compressive}, though others have used Dirichlet \citep{gori2018sigfit, zito2024compressive}, Truncated Normal (L2) \citep{kasar2015whole, kim2016somatic}, or Exponential (L1) priors \citep{kasar2015whole, kim2016somatic}. 

Mutational signature analysis is a rapidly growing field driven by an increasing number of publicly available databases, including COSMIC \citep{tate2019cosmic}, mSignatureDB \citep{huang2018msignaturedb}, PCAWG \citep{icgc2020pan}, and the compendium of Mutational Signatures of Environmental Agents \citep{kucab2019compendium}. With increased data sizes and availability, computational efficiency must be carefully considered when making modeling and implementation choices.

\section{Methodology}\label{sec:methods}

\subsection{MH-within-Gibbs for Computational Efficiency}
\label{sec:methods_mh}

We introduce two efficient models for Poisson Bayesian NMF. The first uses a Truncated Normal prior on each element of $P$ and $E$: 
\begin{align}
    \text{Likelihood: }&\quad M_{kg} \sim \text{Poisson}((PE)_{kg})\\
    \text{Priors: } &\quad P_{kn} \sim \text{TruncNorm}(\mu^P_{kn}, \sigma^{P}_{kn}, 0, \infty), \quad E_{ng} \sim \text{TruncNorm}(\mu^E_{ng}, \sigma^{E}_{ng}, 0, \infty).
\end{align}
Specifications of hyperprior distributions and hyperparameters can be found in Appendix~A.

The target distributions $f$ for Gibbs updates of $P$ and $E$ are not easily sampled. Using $P_{kn}$ as an example, let $P^*_{kn}$ be a potential new value, and let $P^*$ be the $P$ matrix with element $k,n$ replaced by $P^*_{kn}$. Each term $(P^*E)_{kg}$ depends on $P^*_{kn}$ for all $n$, linking likelihoods across $n$, and preventing factorization into a standard family:
\begin{align} \label{eq:target}
    f(P_{kn}^*|P_{-kn}, E, M) &\propto p_{\text{TruncNorm}}(P_{kn}^*)\prod_g p_{\text{Poisson}}\left(M_{kg}|\lambda = (P^*E)_{kg}\right)\\
    &\propto p_{\text{TruncNorm}}(P^*_{kn})e^{-\sum_g (P^*E)_{kg}}\prod_g (P^*E)_{kg}^{M_{kg}}\nonumber.
\end{align}

Poisson augmentation resolves this in previous methods \citep{cemgil2009}, but at high computational cost. Instead, we use Metropolis-Hastings (MH) steps for Gibbs updates of the $P$ and $E$ matrices. We refer to models using this approach as ``Poisson+MH'' from here on. We construct proposal distributions $g$ from the full conditional form of a Normal Bayesian NMF model with identical Truncated Normal priors to our target model:
\begin{align}
    g(P^*_{kn}|P_{-kn}, E, M)& \propto p_{\text{TruncNorm}}(P^*_{kn}|\theta)\prod_g p_{\text{Normal}}\left(M_{kg}|\mu = (P^*E)_{kg}, \sigma^2 = (PE)_{kg}\right) \label{eq:prop} \\
    & = p_\text{TruncNorm}\left(P^*_{kn} | \mu = m(M,P,E,\theta), \sigma = s(M,P,E,\theta), 0, \infty\right) \nonumber
\end{align}
where $\theta$ holds prior and hyperprior parameters. The Normal model is never sampled directly. Specifications of the location $m(\cdot)$ and scale $s(\cdot)$ functions are in Appendix A.

Instead of modeling variances $\sigma^2_{kg}$ as would be required in a Normal Bayesian NMF, we substitute $(PE)_{kg}$ into the proposal. This allows the proposal to reflect the Poisson mean-variance relationship of the target model without additional parameters. This affects only the proposal $g$, which needs to be a proper distribution in order to maintain detailed balance, but does not need to be an actual full conditional.

After drawing a proposed value $P^*_{kn}$, the acceptance ratio can be computed as a ratio of likelihoods because priors cancel out from both the target and proposal distributions:
\begin{align}
    a^P_{kn} &= \frac{f(P^*_{kn}|P_{-kn}, E, M, \theta)g(P_{kn}|P^*, E, M, \theta)}{f(P_{kn}|P_{-kn}, E, M, \theta)g(P^*_{kn}|P, E, M, \theta)} \\
    &= \prod_g\frac{p_{\text{Poisson}}\left(M_{kg}|\lambda = (P^*E)_{kg}\right) \cdot p_{\text{Normal}}\left(M_{kg}|\mu = (PE)_{kg}, \sigma^2 = (P^*E)_{kg}\right)}{p_{\text{Poisson}}\left(M_{kg}|\lambda = (PE)_{kg}\right) \cdot p_{\text{Normal}}\left(M_{kg}|\mu = (P^*E)_{kg}, \sigma^2 = (PE)_{kg}\right)}.
\end{align}

For a fixed rank, the Normal and Poisson Bayesian NMF models have approximately the same MAP solutions (up to scaling and permutation symmetries of NMF \citep{lee1999learning, donoho2003does}). Consequently, the Normal-based proposal captures the local curvature of the Poisson posterior near its mode, designed to have high overlap.

The second model we introduce places an Exponential prior on elements of $P$ and $E$. In order to adapt the Poisson+MH approach to this model, the proposal distribution is now based on a paired Normal likelihood NMF with Exponential priors. This proposal again takes the form of a Truncated Normal distribution but with redefined functions $m(\cdot)$ and $s(\cdot)$. 

In general, the Poisson+MH approach can only be used when the prior (which we assume is shared by the target and the proposal) leads to Normal Bayesian NMF full conditionals that are of known families, meaning Gamma priors cannot be used with this setup. The limitation is practical rather than fundamental---other proposal families could, in principle, accommodate Gamma priors but are not considered here.
  
\subsection{Sparse Bayesian Factor Inclusion (SBFI) for Learning Rank}\label{sec:learning-rank-methods}

Previous Bayesian work by \citet{grabski2023bayesianFA, grabski2025bayesian} in the multi-study setting learns factor sharing patterns and total latent dimension using a binary factor inclusion matrix $A$. We adapt this approach to learn rank automatically in our single-study setting and introduce a new prior and hyperprior structure such that the induced prior on the expected rank is approximately uniform. We refer to this as Bayesian factor inclusion (BFI).

BFI uses a diagonal binary factor inclusion matrix, $A$, where $A_{nn} = 1$ if latent factor $n$ is included in the model, and $0$ if excluded, redefining NMF as
\begin{align*}
   \mathbb{E}[M_{kg}] = \sum_{n = 1}^{N} P_{kn}A_{nn}E_{ng}, \quad\quad \mbox{i.e.,} \quad \quad \mathbb{E}[M] = PAE
\end{align*}
with a prespecified large $N$ and learned latent rank $N' = \sum_{n = 1}^{N}A_{nn}$. If $A$ is fixed as an identity matrix, this formulation is identical to that in Section \ref{sec:methods_mh} for latent rank $N$.

Each factor inclusion indicator $A_{nn}$ has a Bernoulli prior. The prior probability of factor inclusion $q$ is determined by a random variable $R$ representing the expected rank with a uniform hyperprior. The value $q$ is bounded away from 0 and 1 to prevent the Gibbs sampler from getting stuck at the endpoint ranks 0 or $N$:
\begin{align*}
    &A_{nn} \sim \text{Bernoulli}(q),\quad\quad q = R/N \text{, truncated to }q\in \{0.4/N, 1-0.4/N\}\\
    &p(R = r) = 1/(N + 1) \text{ for } r = 0,..., N.
\end{align*}
If  $A_{nn}$ is sampled as $0$, hyperpriors of $P$ and $E$ retain some information on the excluded signature, reducing the extent of label switching.

Including $A$ in the model induces a highly multimodal posterior. We adapt the tempering approach used in \citet{grabski2025bayesian} for the Gibbs updates of $A$ and $R$, where the likelihood term is raised to a temperature power $\gamma$: $A_{nn}\sim p(A_{nn})p(M|A_{nn}, ...)^\gamma$ and $R \sim p(R)p(M|R, ...)^\gamma$. The temperature $\gamma$ progresses from 0 to 1 over the course of a burn-in period. Early samples are from the prior alone, allowing better exploration of the parameter space, and once $\gamma = 1$, samples are from the full conditional. No tempered samples are used for posterior inference.

We also introduce Sparse BFI (SBFI) with a sparse prior on $A_{nn}$:
\begin{align}
    \label{sbfi_def}
    p&_{\small \text{SBFI}}(A_{nn} = 1) = \frac{qG^{-\frac{1}{2}(K+ G)}}{qG^{-\frac{1}{2}(K+ G)} + (1-q)}\nonumber \\
    p&_{\small \text{SBFI}}(A_{nn} = a|...) \propto q^a(1-q)^{1-a} \cdot \left[p(M|A_{nn} = a, ...) \cdot G^{-\frac{1}{2}(K + G)N^{(a)}}\right]^\gamma
\end{align}
where $G$ is sample size, $N^{(a)} = \sum_{n'}A_{n'n'}$ when $A_{nn} = a$, and $(K + G)N^{(a)}$ is the number of functional parameters in $P$ and $E$. During the tempering phase of the sampler, tempering is also applied to the sparsity-inducing term to avoid early convergence on a low rank before signatures are learned. Again, no tempered samples are used for posterior inference.

The structure of this prior results in a BIC-like penalty on rank. Specifically, once tempering is complete, the Gibbs update in Equation \ref{sbfi_def} is identical to the product of the standard BFI prior and the term $e^{-\frac{1}{2} \text{BIC}}$:
\begin{align*}
    &\text{BIC} = -2\cdot \text{log} \left[p_{\text{Poisson}}(M | A_{nn} = a, ...)\right] + \text{log} (G) \cdot (K+G)N^{(a)},\\
    &q^a(1-q)^{1-a} \cdot e^{-\frac{1}{2} \text{BIC}} = q^a(1-q)^{1-a} \cdot p(M|A_{nn} = a, ...) \cdot G^{-\frac{1}{2}(K + G)N^{(a)}}.
\end{align*}

\subsection{Baseline Models}
For comparison, we implement standard Gibbs samplers for (i) Poisson Bayesian NMF with Exponential and Gamma priors (using Poisson augmentation), and (ii) Normal Bayesian NMF with Exponential and Truncated Normal priors. Standard Poisson models serve as direct baselines, against which our novel Poisson+MH models reach equivalent solutions at reduced computational cost. The Normal models provide motivation and validation of our Normal model-based proposals within the Poisson+MH framework, as they converge to the same modes as the standard Poisson models. All models are summarized in Table~\ref{tab:models}.

The Poisson-Gamma model is available in {\tt SigneR} \citep{rosales2016} while Truncated Normal and Exponential priors as well as Normal likelihoods are available in {\tt SignatureAnalyzer} \citep{kasar2015whole, kim2016somatic}. We implement our own samplers to enable direct model comparisons where all settings but likelihoods and priors are held constant. We do not implement a standard Gibbs sampler for the Poisson-Truncated Normal model because the full conditionals are not from distributions that can be directly sampled, even with Poisson augmentation. The Poisson-Gamma model is the closest comparison.

To evaluate the ability of SBFI to learn rank, we compare it to minBIC, a procedure that fits a separate Poisson+MH sampler for each candidate rank and selects the one minimizing BIC. We also compare it to automatic relevance determination as implemented in {\tt SignatureAnalyzer}, which learns rank but does not report posterior uncertainty. Under the non-negativity constraints in NMF, SignatureAnalyzer's L1 and L2 priors correspond to Exponential and Truncated Normal priors, respectively.

\begin{table}[]
    \centering
    \begin{tabular}{|l|l|l|l|}
        \hline
        \textbf{Likelihood} & \textbf{Prior} & \textbf{Sampler}  & \textbf{Purpose}\\
        \hline
        Poisson & Truncated Normal & MH-within-Gibbs & This Work \\
        Poisson & Exponential      & MH-within-Gibbs & This Work \\
        Poisson & Gamma            & Standard Gibbs  & Baseline  \\
        Poisson & Exponential      & Standard Gibbs  & Baseline \\
        Normal  & Truncated Normal & Standard Gibbs  & Motivation \\
        Normal  & Exponential      & Standard Gibbs  & Motivation \\
        \hline
    \end{tabular}
    \caption{List of models implemented in {\tt bayesNMF}. MH-within-Gibbs samplers are the new contribution of this work. Standard Poisson samplers serve as a baseline for comparison to validate MH-within-Gibbs models. Standard Normal samplers serve as motivation for the proposal distributions of MH-within-Gibbs models.}
    \label{tab:models}
\end{table}

\subsection{Inference} \label{sec:inference}

For results in Sections \ref{sec:simulations} and \ref{sec:data}, inference is performed on the final 1000 posterior samples, though our software allows this number to be user-specified and for all samples to be stored.

To summarize the posterior distribution when rank is fixed, we first remove the scale indeterminacy of NMF by normalizing each iteration's $P^{(i)}$ and $E^{(i)}$ matrices so that columns of $P^{(i)}$ sum to one. Then, posterior means $\hat P$ and $\hat E$ are element-wise averages, and 95\% credible intervals are element-wise 2.5th and 97.5th percentiles. 

Parameters $P$ and $E$ are only interpretable in the context of the inclusion structure $A$. Therefore, when learning rank, the target for inference is the posterior distribution conditional on the maximum a-posteriori (MAP) inclusion matrix $\hat A$, identified as the mode across samples $A^{(i)}$. Summaries of the conditional posterior are then computed from samples where $A^{(i)} = \hat A$. Within the selected samples, $\hat P$ and $\hat E$ are computed as above.

For optional post-hoc comparison with a reference $P$, we assign each $P^{(i)}$ to $P$ with the Hungarian algorithm \citep{kuhn1955hungarian}. This assumes each reference signature may only be assigned to a single estimated signature, and vice versa. We ensemble assignments across samples with majority voting (details in Appendix~B.4). 

\section{Implementation and R Software Package} \label{sec:software}

\subsection{Software}

The {\tt bayesNMF} R software package is available for download on GitHub at \href{https://github.com/jennalandy/bayesNMF}{jennalandy/bayesNMF}. Details of model specifications, Gibbs updates, and hyperparameters are available in Appendix A. Details of automated convergence detection and diagnostics, inference, and reference assignment are in Appendix B. 

\subsection{Modeling Capabilities}

Users are able to fit all models described in this paper with the {\tt bayesNMF} function. Model specifications can be adjusted by the {\tt likelihood}, {\tt prior}, and {\tt MH} parameters. The {\tt rank} can be a fixed value or a range vector, in which case {\tt rank\_method} specifies whether minBIC, SBFI, or BFI is used to learn rank. Users are also able to set hyperprior parameters or specify initial values of parameters or prior parameters.

\subsection{Reference Comparison and Visualization Capabilities}

\begin{figure}
    \centering
    \includegraphics[width=\linewidth]{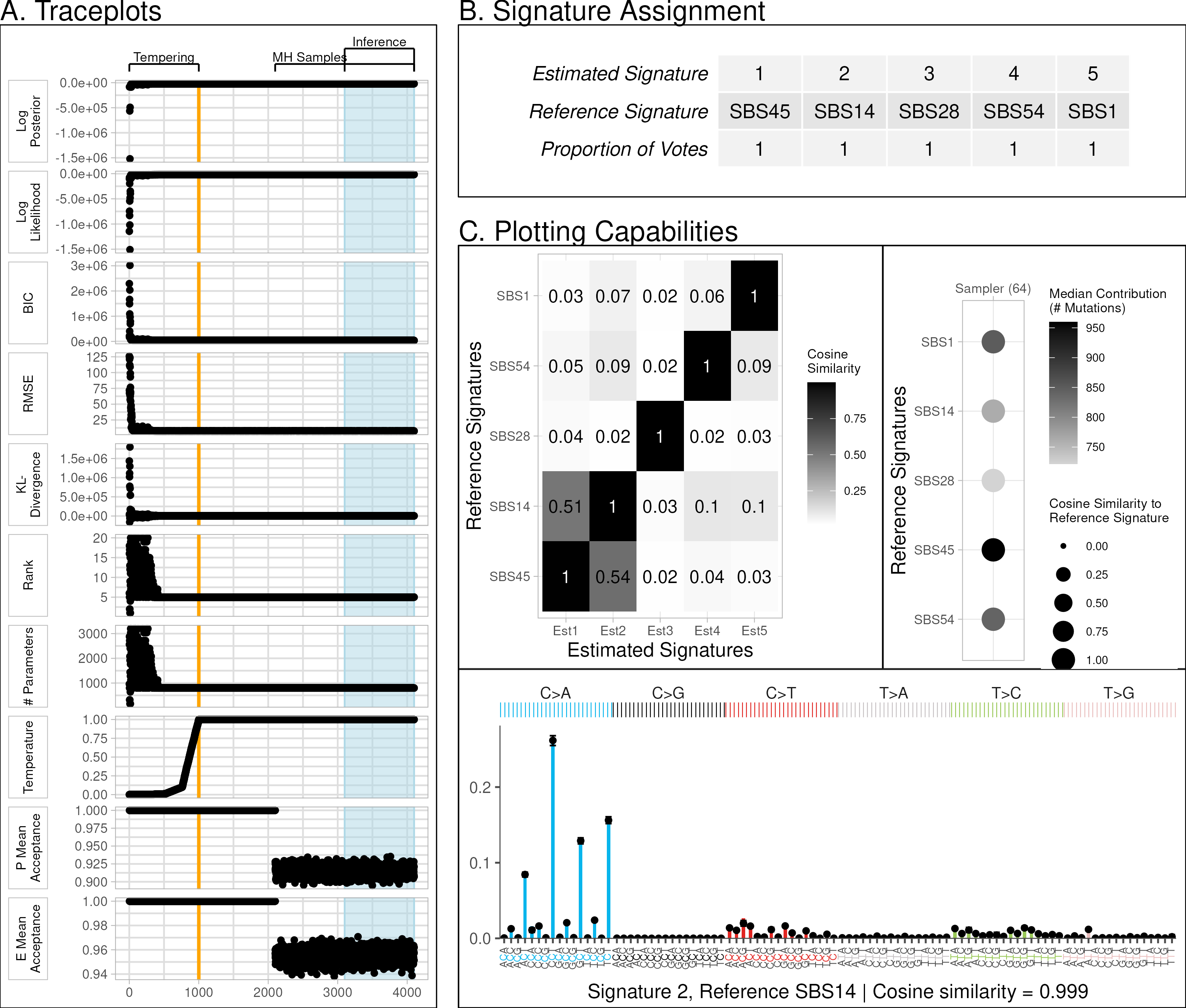}
    \caption{Illustration of software package capabilities using {\tt bayesNMF} Poisson-Truncated Normal+MH SBFI on simulated data. \textbf{A.} Posterior diagnostic traceplots. \textbf{B.}~Reference assignment using posterior ensemble with majority voting. \textbf{C.} Visualization suite, including similarity heatmaps, contribution summaries, and reconstructed signatures (bar chart of aligned reference, points for final estimates, and error bars for 95\% credible intervals).}
    \label{fig:capabilities}
\end{figure}

The {\tt bayesNMF} R package allows users to visualize results and optionally compare them to a set of reference signatures (the default are COSMIC v3.3.1 SBS signatures \citep{tate2019cosmic}). Figure \ref{fig:capabilities}B provides an example output of the reference assignment.

The {\tt plot} function creates multiple visualizations comparing the sampler results to the reference. Figure \ref{fig:capabilities}C shows a cosine similarity heatmap between estimated signatures and assigned references, a dot plot highlighting the median number of mutations attributed to each signature and posterior mean cosine similarity, as well as estimated mutational signatures with posterior uncertainty. Additional functions, plot variations, and a visual diagnostic for label switching are documented in the package \href{https://github.com/jennalandy/bayesNMF/blob/master/vignettes/}{vignettes} and in Appendix~B.5.

\begin{figure}
    \centering
    \includegraphics[width = \linewidth]{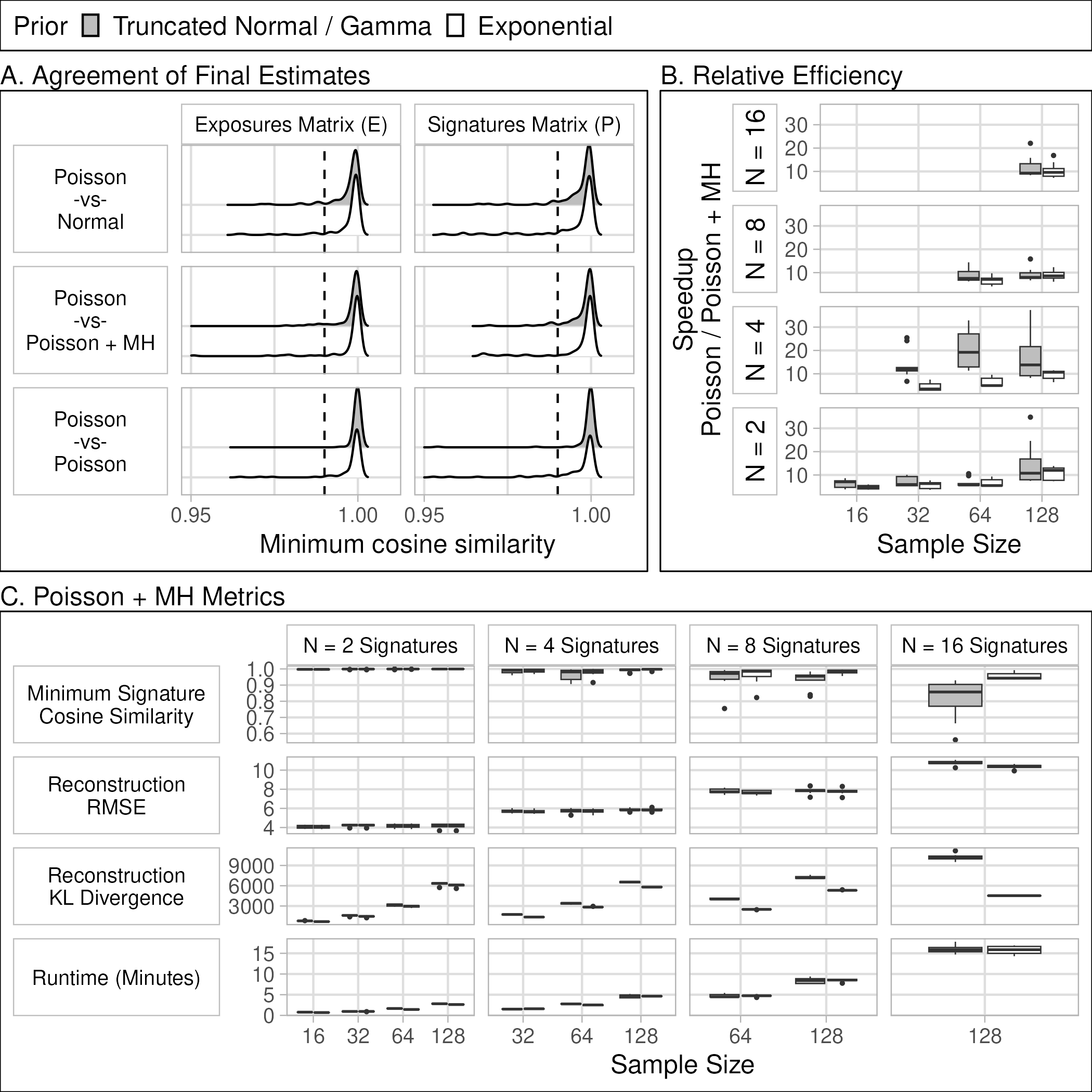}

    \caption{Relative performance of Bayesian NMF models. Grey fill represents Gamma priors on standard Poisson models, but Truncated Normal priors on Poisson+MH or Normal models. White fill always indicates Exponential priors. 
    \textbf{A.} Agreement of $\hat E$ (left) and $\hat P$ (right) between models as signature-wise minimum cosine similarities. 
    Excluding 0 to 7 values per density for similarity $<$0.95 for visual clarity (see Appendix Table C.1).
    \textbf{B.} Efficiency gain of Poisson+MH relative to standard Poisson. 
    \textbf{C.} Performance of novel Poisson+MH models. Metrics for all models are available in Appendix Figure C.1.}
    \label{fig:fixed_rank_results}
\end{figure}

\section{Simulation Studies} \label{sec:simulations}

\subsection{Metrics}

Before computing metrics, signatures are aligned to ground truth. Results for bayesNMF are aligned as described in Section \ref{sec:inference}. SignatureAnalyzer does not report posterior samples, so its results are aligned using point estimates only.

When rank is known, and two models are compared to one another, we report the cosine similarity of the worst-aligned signature, as well as that of the worst-aligned exposure profile. If the two solutions are identical, both values will be exactly 1. We also report relative efficiency as the ratio of total runtime between two models.

When rank is known, and a model is compared to ground truth, we first evaluate NMF solutions in terms of their ability to reconstruct the original data matrix using RMSE and KL-Divergence between $M$ and $\hat M = \hat P \hat E$. We then measure their ability to reconstruct ground truth latent factors using the cosine similarity of the worst-reconstructed signature. Finally, we report the total runtime in minutes to assess computational efficiency.

When learning rank, we evaluate NMF solutions by the bias of the estimated rank as well as post-alignment sensitivity and precision. Sensitivity is the proportion of true signatures that have an estimated match of cosine similarity $>$0.9. It is penalized by underestimating rank because at least one true signature will be without an estimated match. Precision is the proportion of estimated signatures that have a true match. It is penalized by overestimating rank. It may also be penalized by underestimating rank if estimated signatures are combinations of true signatures, but not if one or more true signatures are simply excluded.

\subsection{Poisson+MH efficiently samples the correct posterior} \label{sec:fixed-rank-sim}

In this section, we treat rank as known and perform model comparisons to further justify our Normal model-based proposals and to show that Poisson+MH models more efficiently sample the same posterior as standard Poisson models.

We consider true latent ranks $N \in [2,4,8,16]$ and sample sizes $G \in [16,32,64,128]$. For each $N,G$ where $G \ge 8 N$, 10 datasets are generated. Signatures are sampled from COSMIC SBS v3.3.1 \citep{tate2019cosmic}, each normalized to sum to 1. The expected total mutation count for sample $g$ is ${m_g \sim \text{Negative Binomial}(111.11 N, 0.1)}$, yielding $E[m_g] = 1000  N$, or $1000$ mutations per signature per sample. This gives each simulation comparable relative power to learn each signature, and this explains why reconstruction errors increase with $N$. Columns of the exposures matrix, $E$, are generated $E_g \sim \text{Multinomial}(m_g, \mathbf p_g)$, where $\mathbf p_g \sim \text{Dirichlet}(\mathbf{1}_N)$. Mutation counts are simulated as $M_{kg} \sim \text{Poisson}((PE)_{kg})$.

Normal Bayesian NMF results in point estimates nearly identical to Poisson Bayesian NMF, justifying our Normal model-based MH proposals (Figure \ref{fig:fixed_rank_results}A, top row). Further, the addition of MH steps does not alter point estimates of Poisson Bayesian NMF (middle row). These comparisons yield agreement $>$0.99 in at least 84\% of cases and are very close to the null case comparing two independent chains of the standard Poisson sampler (bottom row). Along with aligned point estimates, the addition of MH steps increases effective sample sizes (Appendix C.1) and does not substantially change credible interval widths (Appendix C.3). Together, these results show that our novel Poisson+MH samplers yield the same posterior inference as the standard Poisson samplers. Additionally, Poisson+MH models are 3-30x faster, and this gap grows with dimensionality (Figure \ref{fig:fixed_rank_results}B).

\subsection{SBFI learns rank}\label{sec:sim_learned}

In this section, we demonstrate that SBFI performs comparably to or better than SignatureAnalyzer's automatic relevance determination (ARD) while reporting posterior uncertainty.

\begin{figure}
    \centering
    \includegraphics[width = \linewidth]{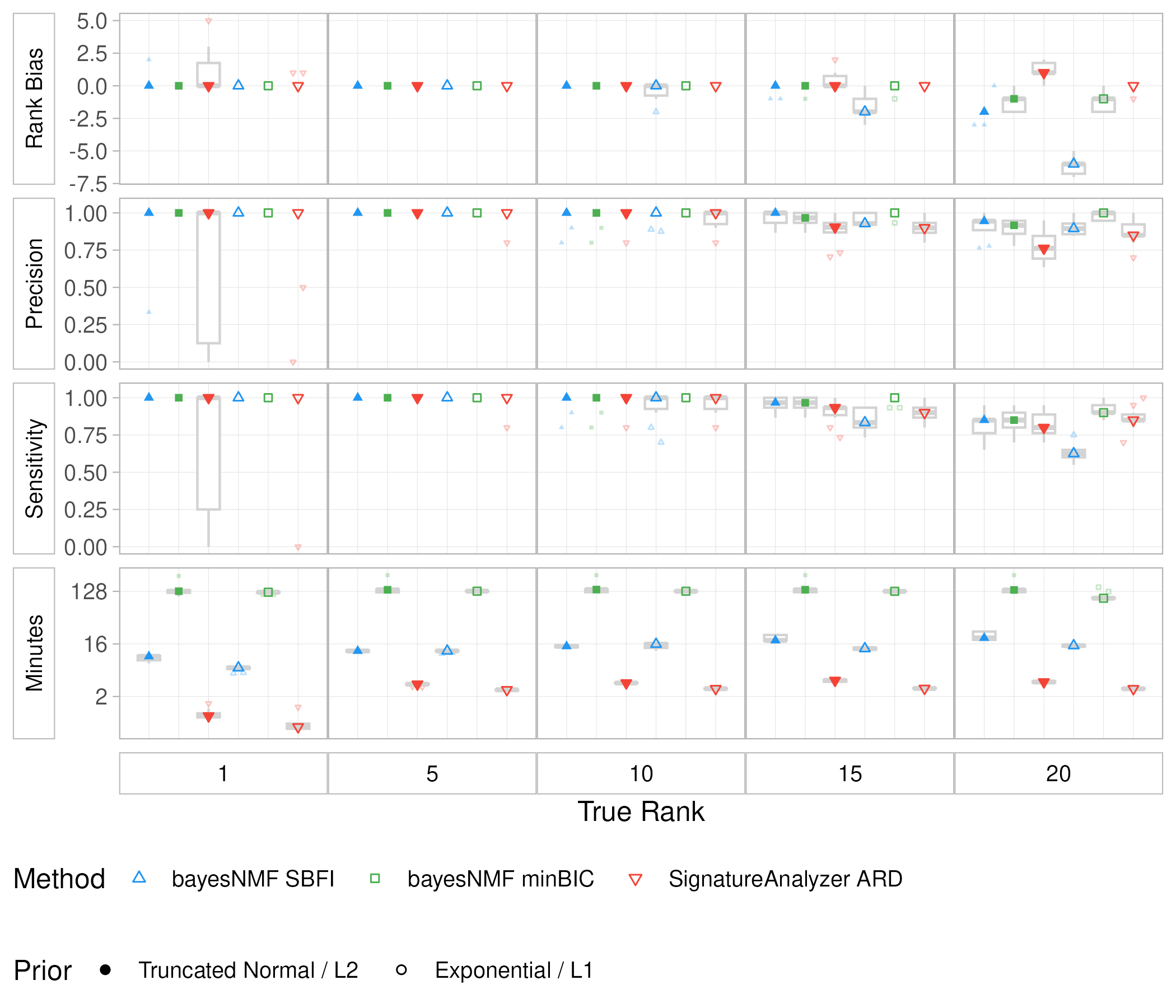}
    \caption{Rank bias, precision, sensitivity, and time (log scale) of rank learning approaches: bayesNMF with SBFI (upwards triangle) and minBIC (square), as well as SignatureAnalyzer with ARD (downwards triangle), each with Truncated Normal (filled) and Exponential (empty) priors. Plotted are large shapes for medians on top of lighter boxplots and jittered outliers. Full results for all ranks between 1 and 20 are available in Appendix Figure C.3. Sensitivity is the proportion of true signatures for which there is an estimated signature with cosine similarity $>$0.9. Precision is the proportion of estimated signatures for which there is a true signature with cosine similarity $>$0.9.}
    \label{fig:learn-rank-res}
\end{figure}

Data are simulated as in Section \ref{sec:fixed-rank-sim} with $G = 64$ and 10 datasets for each true rank $N$ between $1$ and $20$. For each dataset, no two signatures have a cosine similarity $\ge$0.8. Reduced precision and sensitivity observed at larger ranks likely reflect the difficulty of the problem (i.e., large $N$ relative to $G$), rather than deficiencies of the methods considered.

Figure \ref{fig:learn-rank-res} reports rank bias, precision, sensitivity, and runtime for each model as true rank increases along the x-axis. For a Truncated Normal / L2 prior (filled shapes), all approaches estimate rank well, but with different error patterns: SignatureAnalyzer overestimates low and high ranks, while SBFI and minBIC underestimate high ranks. Under this prior and at high ranks, SBFI and minBIC achieve the highest precision and sensitivity, meaning that the signatures our models do estimate are correct, even though some are missed, while SignatureAnalyzer often yields noisy or split signatures, even if it learns the correct rank.  For an Exponential / L1 prior (empty shapes), the sparsity penalty of SBFI may be too strong, as SBFI underestimates high ranks by a wide margin. In this setting, minBIC still shows higher precision and sensitivity than SignatureAnalyzer.

With our Poisson+MH samplers, we are able to report posterior uncertainty at the cost of 12 additional minutes over SignatureAnalyzer (4 vs 16 minutes). This would take multiple hours using SBFI with standard Poisson samplers. Although choosing rank by optimizing BIC is not a new idea, our Poisson+MH samplers also allow the minBIC strategy to complete much faster: the minBIC approach takes about two hours with Poisson+MH samplers, but would take over 20 hours with standard Poisson samplers.

\section{Analysis of SBS Signatures in PCAWG Data}\label{sec:data}

\subsection{Data}

In this section, we analyze the Pan-Cancer Analysis of Whole Genomes (PCAWG) mutation count data using the 96-alphabet single base substitution (SBS) mutation classification \citep{icgc2020pan}. This is publicly available from the International Cancer Genome Consortium’s (ICGC) Accelerating Research in Genomic Oncology (ARGO) data portal \citep{zhang2019international}. This includes 37 datasets, each for a different histology group. We excluded histology groups with 10 or fewer samples, as well as benign (non-cancerous) histology groups, leaving 32 datasets for analysis.

We exclude hypermutated samples from all analyses. The inclusion of hypermutated samples can result in \textit{signature bleeding}, where mutations in non-hypermutated samples are misattributed to hypermutation-specific signatures \citep{maura2019practical, alexandrov2020repertoire}. For each histology group, we identify hypermutated samples by assuming the total mutation count per sample follows a Negative Binomial mixture model, with at least one component for non-hypermutated samples, and possibly one or more components for hypermutated samples (details in Appendix D.1). This removes 65 samples total ($<$3\% of data), with at most 13 samples removed per histology group in skin melanoma.

\subsection{Methods}\label{sec:pcawg_methods}

Because ground truth is unknown for real data applications, rigorous conclusions can only be made in the context of simulated data, as in Section \ref{sec:simulations}. However, we wish to illustrate our method and compare it to others in a real-world setting. We compare our results to SignatureAnalyzer and show that our approach results in similar biological interpretations.

We use the bayesNMF Poisson+MH approach with Truncated Normal priors and SBFI for rank $\in [1,20]$. For each histology group, we fit bayesNMF+MH, and compare to SignatureAnalyzer with L2 priors.

\begin{figure}
    \centering
    \includegraphics[width = \linewidth]{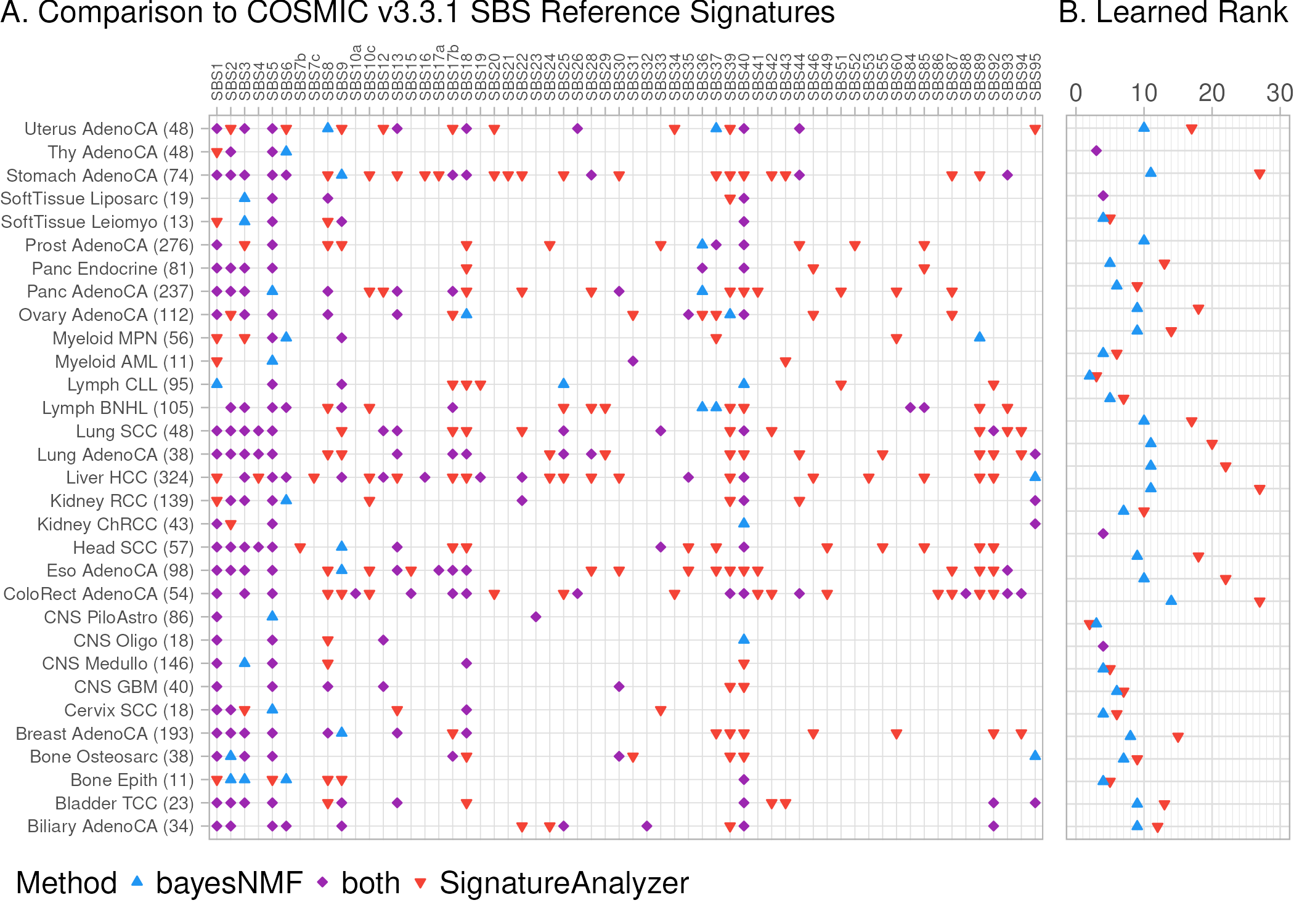}
    \caption{Results of bayesNMF+SBFI (upwards triangle) and SignatureAnalyzer+ARD (downwards triangle) on PCAWG histology groups. Skin melanoma is excluded for visual clarity because SignatureAnalyzer estimates a high rank of 76 (bayesNMF estimates a rank of 10). \textbf{A.} COSMIC reference signatures aligned to $\hat P$. Reference signatures not found with either method are excluded. \textbf{B.} Estimated latent ranks. Exact values in Appendix Table D.1.}
    \label{fig:pcawg_res}
\end{figure}

\subsection{Results}

Figure \ref{fig:pcawg_res}A shows a high degree of overlap in the signatures discovered by bayesNMF+SBFI and SignatureAnalyzer+ARD. These results mirror many trends from \citet{alexandrov2020repertoire}, including the presence of SBS1, SBS5, and SBS40 in nearly every cancer type. More detailed results, including the median number of mutations attributed to each signature in each histology group by each method, can be found in Appendix Figure D.2.

A clear difference between methods is the estimated rank---bayesNMF+SBFI typically results in a sparser solution, with SignatureAnalyzer+ARD frequently estimating twice the number of signatures (Figure \ref{fig:pcawg_res}B). Based on our simulation results, because bayesNMF+SBFI with Truncated Normal priors consistently achieved higher sensitivity and precision, we speculate that the results here of SignatureAnalyzer with L2 priors may include spurious components (especially for skin melanoma, where SignatureAnalyzer estimates a rank of 76), making the resulting signatures less biologically interpretable.

In most cases where bayesNMF recovered a signature not reported by SignatureAnalyzer (Figure \ref{fig:pcawg_res}A upwards triangles), the discrepancy can be partly explained by posterior uncertainty. These signatures have lower cosine similarity with their assigned reference signatures and a lower proportion of assignment votes, indicating uncertainty in assignment (Appendix Figure D.3). However, of the 34 signatures of this type, 28 (82\%) had a SignatureAnalyzer-derived signature among the candidates receiving assignment votes, just not the majority required for posterior ensemble assignment. Therefore, many signatures estimated by SignatureAnalyzer can be viewed as consistent with one plausible posterior mode inferred by bayesNMF. Still, there are 6 signatures unique to bayesNMF after considering this aspect, of which many match previous discoveries in the COSMIC database: SBS5 in CNS pilocytic astrocytoma and pancreatic adenocarcinoma, SBS1 in lymphocytic leukemia (CLL), and SBS95 in liver hepatocellular carcinoma (HCC). Some are not in the COSMIC database, but have been previously reported in the literature: SBS36 in prostate adenocarcinoma \citep{jiang2020global}. There is no reference for SBS9 in esophageal adenocarcinoma, although it has been found in esophageal small cell carcinoma.

\section{Discussion}

We introduced bayesNMF, a MH-within-Gibbs sampler for Bayesian Poisson NMF with automatic rank-learning. The method is built on two primary innovations: an efficient MH-within-Gibbs sampling strategy to avoid Poisson augmentation, and a BIC-based sparsity prior that enables rank to be learned accurately within a single Bayesian formulation. Simulation studies and a large-scale data analysis demonstrate that bayesNMF achieves comparable accuracy and uncertainty quantification to standard Bayesian Poisson NMF at a fraction of the computational cost. All methods are available in an open-source R software package on GitHub at \href{https://github.com/jennalandy/bayesNMF}{jennalandy/bayesNMF}.

The MH steps with our high-overlap proposals are the key to bayesNMF's computational efficiency. In future work, these updates could be used in conjunction with other approaches such as compressive NMF \citep{zito2024compressive} or multistudy NMF \citep{grabski2025bayesian}. Additionally, our model-paired proposals may be generalizable to other hierarchical count models that rely on Poisson augmentation, like Poisson regression with random effects.

We recognize a potential limitation of our assumption that each estimated signature must be assigned to a unique reference signature. In some cases, the precision reported in Figure \ref{fig:learn-rank-res} could be higher if multiple estimated signatures were allowed to map to the same reference. Similarly, some SignatureAnalyzer-estimated signatures in the PCAWG analysis (Appendix Figure~D.2) may have achieved higher similarity to their assigned references if duplicate assignments were permitted. However, this is only a limitation of our downstream analysis and visualization tools, not of the models themselves.

In summary, bayesNMF offers a powerful toolbox for exploring complex mutational landscapes, especially with its ability to report posterior uncertainty while learning rank as part of the Bayesian model. The open source R software package facilitates the usability of these methods and the reproducibility of our results. As cancer genomic databases expand, bayesNMF's efficiency and flexibility make it well-suited to drive further research.

\section{Acknowledgments}
The authors gratefully acknowledge that this work was supported by the NIH-NIGMS training grant T32GM135117, NIH-NCI grant 5R01 CA262710-03, and NSF-DMS grant 2113707.

\bibliography{references_etal}

@article{grabski2023bayesianFA,
  title={Bayesian combinatorial MultiStudy factor analysis},
  author={Grabski, Isabella N and De Vito, Roberta and Trippa, Lorenzo and Parmigiani, Giovanni},
  journal={The Annals of Applied Statistics},
  volume={17},
  number={3},
  pages={2212},
  year={2023},
  publisher={NIH Public Access}
}

@article{grabski2025bayesian,
  title={Bayesian multi-study non-negative matrix factorization for mutational signatures},
  author={Grabski, Isabella N and Trippa, Lorenzo and Parmigiani, Giovanni},
  journal={Genome Biology},
  volume={26},
  number={1},
  pages={98},
  year={2025},
  publisher={Springer}
}

@article{dempster1977maximum,
  title={Maximum likelihood from incomplete data via the EM algorithm},
  author={Dempster, Arthur P and Laird, Nan M and Rubin, Donald B},
  journal={Journal of the royal statistical society: series B (methodological)},
  volume={39},
  number={1},
  pages={1--22},
  year={1977},
  publisher={Wiley Online Library}
}

@article{zamborszky2017loss,
  title={Loss of BRCA1 or BRCA2 markedly increases the rate of base substitution mutagenesis and has distinct effects on genomic deletions},
  author={Z{\'a}mborszky, Judit and Szikriszt, Bernadett and Gervai, Judit Zsuzsanna and Pipek, Orsolya and P{\'o}ti, {\'A}d{\'a}m and Krzystanek, Marcin and Ribli, Dezs{\H{o}} and Szalai-Gindl, J{\'a}nos M{\'a}rk and Csabai, Istv{\'a}n and Szallasi, Zolt{\'a}n and others},
  journal={Oncogene},
  volume={36},
  number={6},
  pages={746--755},
  year={2017},
  publisher={Nature Publishing Group}
}

@article{kuhn1955hungarian,
  title={The Hungarian method for the assignment problem},
  author={Kuhn, Harold W},
  journal={Naval Research Logistics Quarterly},
  volume={2},
  number={1-2},
  pages={83--97},
  year={1955},
  publisher={Wiley Online Library}
}

@article{hayward2017whole,
  title={Whole-genome landscapes of major melanoma subtypes},
  author={Hayward, Nicholas K and Wilmott, James S and Waddell, Nicola and others},
  journal={Nature},
  volume={545},
  number={7653},
  pages={175--180},
  year={2017},
  publisher={Nature Publishing Group UK London}
}

@article{saini2016impact,
  title={The impact of environmental and endogenous damage on somatic mutation load in human skin fibroblasts},
  author={Saini, Natalie and Roberts, Steven A and Klimczak, Leszek J and others},
  journal={PLoS Genetics},
  volume={12},
  number={10},
  pages={e1006385},
  year={2016},
  publisher={Public Library of Science San Francisco, CA USA}
}

@article{alexandrov2013signatures,
  title={Signatures of mutational processes in human cancer},
  author={Alexandrov, Ludmil B and Nik-Zainal, Serena and Wedge, David C and others},
  journal={Nature},
  volume={500},
  number={7463},
  pages={415--421},
  year={2013},
  publisher={Nature Publishing Group UK London}
}

@article{zito2024compressive,
  title={Compressive Bayesian non-negative matrix factorization for mutational signatures analysis},
  author={Zito, Alessandro and Miller, Jeffrey W},
  journal={arXiv preprint arXiv:2404.10974},
  year={2024}
}

@article{icgc2020pan,
  title={Pan-cancer analysis of whole genomes},
  journal={Nature},
  author = {Consortium, The ICGC/TCGA Pan-Cancer Analysis of Whole Genomes},
  volume={578},
  number={7793},
  pages={82--93},
  year={2020},
  publisher={Nature Publishing Group UK London}
}

@article{nik2015genome,
  title={The genome as a record of environmental exposure},
  author={Nik-Zainal, Serena and Kucab, Jill E and Morganella, Sandro and others},
  journal={Mutagenesis},
  volume={30},
  number={6},
  pages={763--770},
  year={2015},
  publisher={Oxford University Press UK}
}

@article{weinstein2013cancer,
  title={The cancer genome atlas pan-cancer analysis project},
  author={Weinstein, John N and Collisson, Eric A and Mills, Gordon B and others},
  journal={Nature Genetics},
  volume={45},
  number={10},
  pages={1113--1120},
  year={2013},
  publisher={Nature Publishing Group}
}

@article{international2010international,
  title={International network of cancer genome projects},
  author={The International Cancer Genome Consortium},
  journal={Nature},
  volume={464},
  number={7291},
  pages={993},
  year={2010},
  publisher={NIH Public Access}
}

@article{paatero1994positive,
  title={Positive matrix factorization: A non-negative factor model with optimal utilization of error estimates of data values},
  author={Paatero, Pentti and Tapper, Unto},
  journal={Environmetrics},
  volume={5},
  number={2},
  pages={111--126},
  year={1994},
  publisher={Wiley Online Library}
}

@article{lee1999learning,
  title={Learning the parts of objects by non-negative matrix factorization},
  author={Lee, Daniel D and Seung, H Sebastian},
  journal={Nature},
  volume={401},
  number={6755},
  pages={788--791},
  year={1999},
  publisher={Nature Publishing Group UK London}
}

@article{lee2000algorithms,
  title={Algorithms for non-negative matrix factorization},
  author={Lee, Daniel and Seung, H Sebastian},
  journal={Advances in Neural Information Processing Systems},
  volume={13},
  year={2000}
}

@article{cemgil2009,
	title = {Bayesian Inference for Nonnegative Matrix Factorisation Models},
	author = {Cemgil, Ali Taylan},
	year = {2009},
	date = {2009},
	journal = {Computational Intelligence and Neuroscience},
	pages = {1--17},
	volume = {2009},
	langid = {en}
}

@inbook{schmidt2009,
	title = {Bayesian Non-negative Matrix Factorization},
	author = {Schmidt, Mikkel N. and Winther, Ole and Hansen, Lars Kai},
	year = {2009},
	date = {2009},
	publisher = {Springer Berlin Heidelberg},
	pages = {540--547}
}

@article{rosales2016,
	title = {signeR: an empirical Bayesian approach to mutational signature discovery},
	author = {Rosales, Rafael and Drummond, Rodrigo and Valieris, Renan and others},
	year = {2016},
	month = {09},
	date = {2016-09-01},
	journal = {Bioinformatics},
	pages = {8--16},
	volume = {33},
	number = {1},
	langid = {en}
}

@article{alexandrov2020repertoire,
  title={The repertoire of mutational signatures in human cancer},
  author={Alexandrov, Ludmil B and Kim, Jaegil and Haradhvala, Nicholas J and others},
  journal={Nature},
  volume={578},
  number={7793},
  pages={94--101},
  year={2020},
  publisher={Nature Publishing Group UK London}
}

@article{gori2018sigfit,
  title={sigfit: flexible Bayesian inference of mutational signatures},
  author={Gori, Kevin and Baez-Ortega, Adrian},
  journal={bioRxiv},
  pages={372896},
  year={2018},
  publisher={Cold Spring Harbor Laboratory}
}

@article{kim2016somatic,
  title={Somatic ERCC2 mutations are associated with a distinct genomic signature in urothelial tumors},
  author={Kim, Jaegil and Mouw, Kent W and Polak, Paz and others},
  journal={Nature Genetics},
  volume={48},
  number={6},
  pages={600--606},
  year={2016},
  publisher={Nature Publishing Group US New York}
}

@article{kasar2015whole,
  title={Whole-genome sequencing reveals activation-induced cytidine deaminase signatures during indolent chronic lymphocytic leukaemia evolution},
  author={Kasar, S and Kim, J and Improgo, R and others},
  journal={Nature Communications},
  volume={6},
  number={1},
  pages={8866},
  year={2015},
  publisher={Nature Publishing Group UK London}
}

@article{islam2022uncovering,
  title={Uncovering novel mutational signatures by de novo extraction with SigProfilerExtractor},
  author={Islam, SM Ashiqul and D{\'\i}az-Gay, Marcos and Wu, Yang and others},
  journal={Cell Genomics},
  volume={2},
  number={11},
  year={2022},
  publisher={Elsevier}
}

@article{nik2012life,
  title={The life history of 21 breast cancers},
  author={Nik-Zainal, Serena and Van Loo, Peter and Wedge, David C and others},
  journal={Cell},
  volume={149},
  number={5},
  pages={994--1007},
  year={2012},
  publisher={Elsevier}
}

@article{nikzainal2012_mutproc,
  author  = "Serena Nik-Zainal and Ludmil B. Alexandrov and David C. Wedge and others",
  title   = "Mutational Processes Molding the Genomes of 21 Breast Cancers",
  journal = "Cell",
  year    = 2012,
  volume  = "149",
  pages   = "994--1007",
}

@article{tate2019cosmic,
  title={COSMIC: the catalogue of somatic mutations in cancer},
  author={Tate, John G and Bamford, Sally and Jubb, Harry C and others},
  journal={Nucleic Acids Research},
  volume={47},
  number={D1},
  pages={D941--D947},
  year={2019},
  publisher={Oxford University Press}
}

@article{huang2018msignaturedb,
  title={mSignatureDB: a database for deciphering mutational signatures in human cancers},
  author={Huang, Po-Jung and Chiu, Ling-Ya and Lee, Chi-Ching and others},
  journal={Nucleic acids research},
  volume={46},
  number={D1},
  pages={D964--D970},
  year={2018},
  publisher={Oxford University Press}
}

@article{kucab2019compendium,
  title={A compendium of mutational signatures of environmental agents},
  author={Kucab, Jill E and Zou, Xueqing and Morganella, Sandro and others},
  journal={Cell},
  volume={177},
  number={4},
  pages={821--836},
  year={2019},
  publisher={Elsevier}
}

@article{maura2019practical,
  title={A practical guide for mutational signature analysis in hematological malignancies},
  author={Maura, Francesco and Degasperi, Andrea and Nadeu, Ferran and others},
  journal={Nature Communications},
  volume={10},
  number={1},
  pages={2969},
  year={2019},
  publisher={Nature Publishing Group UK London}
}

@article{zhang2019international,
  title={The international cancer genome consortium data portal},
  author={Zhang, Junjun and Bajari, Rosita and Andric, Dusan and Gerthoffert, Francois and Lepsa, Alexandru and Nahal-Bose, Hardeep and Stein, Lincoln D and Ferretti, Vincent},
  journal={Nature Biotechnology},
  volume={37},
  number={4},
  pages={367--369},
  year={2019},
  publisher={Nature Publishing Group US New York}
}

@article{jiang2020global,
  title={Global autozygosity is associated with cancer risk, mutational signature and prognosis},
  author={Jiang, Limin and Guo, Fei and Tang, Jijun and Leng, Shuguan and Ness, Scott and Ye, Fei and Kang, Huining and Samuels, David C and Guo, Yan},
  journal={Cancers},
  volume={12},
  number={12},
  pages={3646},
  year={2020},
  publisher={MDPI}
}

@article{neal2011mcmc,
  title={{MCMC} using Hamiltonian dynamics},
  author={Neal, Radford M and others},
  journal={Handbook of Markov Chain Monte Carlo},
  volume={2},
  number={11},
  pages={2},
  year={2011},
  publisher={Chapman and Hall/CRC}
}

@article{donoho2003does,
  title={When does non-negative matrix factorization give a correct decomposition into parts?},
  author={Donoho, David and Stodden, Victoria},
  journal={Advances in Neural Information Processing Systems},
  volume={16},
  year={2003}
}

@article{bertolacci2019armspp,
  title={Armspp: Adaptive Rejection Metropolis Sampling (ARMS) via ‘Rcpp’},
  author={Bertolacci, M},
  journal={R package version 0.0},
  volume={2},
  year={2019}
}

@article{betancourt2017conceptual,
  title={A conceptual introduction to Hamiltonian Monte Carlo},
  author={Betancourt, Michael},
  journal={arXiv preprint arXiv:1701.02434},
  year={2017}
}

@article{carpenter2017stan,
  title={Stan: A probabilistic programming language},
  author={Carpenter, Bob and Gelman, Andrew and Hoffman, Matthew D and Lee, Daniel and Goodrich, Ben and Betancourt, Michael and Brubaker, Marcus and Guo, Jiqiang and Li, Peter and Riddell, Allen},
  journal={Journal of Statistical Software},
  volume={76},
  pages={1--32},
  year={2017}
}

\newpage
\begin{appendix}
\renewcommand{\thefigure}{\Alph{section}.\arabic{figure}}
\renewcommand{\thetable}{\Alph{section}.\arabic{table}}
\renewcommand{\shorttitle}{bayesNMF: Supplementary Material}

\begin{center}
    {\LARGE Supplementary Material}
\end{center}

\begin{description}

\item[Appendix:] A) Model specifications (likelihoods, priors, hyperpriors) and Gibbs updates. B) Implementation details (convergence, warmup, inference, reference alignment, label switching diagnostic). C) Additional simulation study results, including sensitivity analyses for sparse data and increasing rank range. D) Additional PCAWG data application results.

\item[R package:] The R package {\tt bayesNMF} is publicly available on GitHub at \href{https://github.com/jennalandy/bayesNMF}{jennalandy/bayesNMF}, documented by the \href{https://github.com/jennalandy/bayesNMF/blob/master/README.md}{README} and \href{https://github.com/jennalandy/bayesNMF/tree/master/vignettes}{vignettes}.

\item[Code:] All code needed to reproduce simulation studies and data analysis is publicly available on GitHub at \newline\href{https://github.com/jennalandy/bayesNMF_PAPER}{jennalandy/bayesNMF\_PAPER}, documented by \href{https://github.com/jennalandy/bayesNMF_PAPER/blob/master/README.md}{README.md}.

\item[Dataset:] The PCAWG data are available through the ICGC ARGO data portal: \newline \href{https://docs.icgc-argo.org/docs/data-access/icgc-25k-data#open-release-data---object-bucket-details}{access instructions at docs.icgc-argo.org/docs/data-access/}. Instructions are also included in the code file \href{https://github.com/jennalandy/bayesNMF_PAPER/blob/master/studies/PCAWG/1_pcawg_data.qmd}{1\_pcawg\_data.qmd}.

\end{description}

\newpage
\section{Appendix A: Model Specifications}
\label{supp_sec:appendix_a}
For each model, we provide a formal model definition as well as the Gibbs updates, separately for parameters and prior parameters. If applicable, MH-within-Gibbs parameter updates are also documented with targets, proposals, acceptance ratios, and MH steps. Some pieces are identical across all models, in which case the shared specification is reported in Section \ref{sec:defRA} and the shared update is reported in Section \ref{sec:updateRA}. Table \ref{tab:suppA_models} summarizes the six models implemented in {\tt bayesNMF} and links to the applicable subsection of this Appendix. 

We introduce all models using the simpler prior on the signature inclusion matrix $A$ corresponding to Bayesian Factor Inclusion (BFI). See the main text for details on the prior of $A$ under Sparse BFI (SBFI). If rank is fixed, $A$ is set as an identity matrix and Gibbs updates of $A$ are skipped.

Throughout this appendix, let $\hat M$ be the reconstruction $\hat P \hat A \hat E$ such that $\hat M_{kg} = (\hat P \hat E)_{kg} = \sum_{n}\hat P_{kn}\hat E_{ng}$. Similarly, let $\hat M_{kg/n}$ be the reconstructed value of $M_{kg}$ after excluding signature $n$: $\hat M_{kg/n} = \sum_{n' \ne n}\hat P_{kn'} \hat E_{n'g}$.

Further, when in reference to a potential new value $P^*_{kn}$, let $P^*$ be the $P$ matrix with element $k,n$ replaced by $P^*_{kn}$. Similarly, when in reference to a potential new value $E^*_{ng}$, let $E^*$ be the $E$ matrix with element $n,g$ replaced by $E^*_{ng}$.

The value $\bar M$ is the mean across all count values in the data matrix $M$, representing the general scale of the data.

\begin{table}[b]
    \centering
    \begin{tabular}{|l|l|l|l|l|l|}
        \hline
        \textbf{Likelihood} & \textbf{Prior} & \textbf{Sampler} & \textbf{\# Updates} & \textbf{\# Acc. Ratios} & \textbf{Sec.}\\
        \hline
        Poisson & TruncNormal & MH-within-Gibbs & $3(KN + NG)$ & $KN + NG$ & \ref{sec:P_TN_MH}\\
        Poisson  & Exponential      & MH-within-Gibbs & $2(KN + NG)$ & $KN + NG$ &  \ref{sec:P_E_MH}\\
        Poisson & Gamma  & Standard Gibbs & $3(KN + NG) + KNG$ & $0$ & \ref{sec:P_G} \\
        Poisson & Exponential & Standard Gibbs & $2(KN + NG) + KNG$ & $0$ & \ref{sec:P_E} \\
        Normal  & TruncNormal & Standard Gibbs  & $3(KN + NG) + G$ & $0$ & \ref{sec:N_TN} \\
        Normal & Exponential & Standard Gibbs & $2(KN + NG) + G$ & $0$ & \ref{sec:N_E} \\
        \hline
    \end{tabular}
    \caption{List of models implemented in {\tt bayesNMF}, reporting the number of parameter/prior parameter updates and the number of acceptance rates computed per iteration of the Gibbs sampler. Both values directly impact compute time and memory consumption. Typically $N << K,G$.}
    \label{tab:suppA_models}
\end{table}

\subsubsection{Definitions Shared Across Models} \label{sec:defRA}

The prior and hyperprior structure on signature inclusion is identical across all models:
{\footnotesize
\begin{align*}
    \text{\underline{Priors}: }& A_{nn} \sim \text{Bernoulli}(q)\\
    & \text{where }q = \begin{cases}0.4/N & \text{for } R = 0 \\
        R/N & \text{for } R = 1,...,N - 1\\
        1-0.4/N & \text{for } R = N\end{cases},\\
    \text{\underline{Hyperpriors}: }&p(R = r) = 1/(N + 1) \text{ for } r = 0,..., N.
\end{align*}
}

\subsubsection{Parameter Gibbs Updates Shared Across Models} \label{sec:updateRA}

Gibbs updates for $R$ and $A$ are identical across all models, with the appropriate likelihood (Poisson or Normal) plugged in for $p(M|...)$:
{\footnotesize
\begin{align*}
    p(R = r|...) &\propto \frac{1}{N+1}\left[\prod_{n = 1}^N\left(q^{(r)}\right)^{A_n}\left(1-q^{(r)}\right)^{1-A_n}\right]^\gamma\\
    &\text{where } q^{(r)} = \begin{cases}
            0.4/N &\text{if }r = 0\\
            r/N &\text{if } r \in \{1,..., N-1\}\\
            1-0.4/N &\text{if }r = N
        \end{cases}\\
    A_{nn}|... &\sim \text{Bernoulli}\left(\frac{p_1}{p_1 + p_0}\right)\\
    &\text {where } p_a = \begin{cases} q^a (1-q)^{1-a} \cdot p\left(M|A_{nn} = a, ...\right)^{\gamma} & \text{ in BFI}\\ q^a (1-q)^{1-a} \cdot \left[p\left(M|A_{nn} = a, ...\right) \cdot G^{-\frac{1}{2}(K + G)N^{(a)}}\right]^{\gamma} & \text{ in SBFI}\end{cases}\\
    &\text{and }q = q^{(R)}
\end{align*}
}

When the latent rank is provided, the tempering parameter $\gamma$ is always 1. Otherwise, $\gamma$ starts at $0$ and gradually reaches $1$ approximately 20\% of the way through sampling. If $\gamma = 0$, the update distribution is equivalent to the prior, and if $\gamma = 1$, it is the true full conditional. No tempered samples are used for inference.

\newpage
\subsection{Poisson - TruncNormal(TN) + MH} \label{sec:P_TN_MH}

\subsubsection{Model Definition}

{\footnotesize
\begin{align*}
    \text{\underline{Likelihood}: }& M_{kg} \sim \text{Poisson}\left(\sum_{n}P_{kn}A_{nn}E_{ng}\right)\\
    \text{\underline{Priors}: }& P_{kn} \sim \text{TruncNorm}\left(\mu^P_{kn}, (\sigma^P_{kn})^2, 0, \infty\right)\\
    & E_{ng} \sim \text{TruncNorm}\left(\mu^E_{ng}, (\sigma^E_{ng})^2, 0, \infty\right)\\
    \text{\underline{Hyperpriors}: }& \mu^P_{kn} \sim \text{Normal}\left(m^P_{kn} = 0, (s^2_{kn})^2 = \sqrt{\bar M/N} \right)\\
    &(\sigma^P_{kn})^2 \sim \text{InverseGamma}\left(a^P_{kn} = N + 1, b^P_{kn} = \sqrt{N}\right)\\
    &\mu^E_{ng} \sim \text{Normal}\left(m^E_{ng} = 0, (s^E_{ng})^2 = \sqrt{\bar M/N} \right)\\
    &(\sigma^E_{ng})^2 \sim \text{InverseGamma}\left(a^E_{ng} = N + 1, b^E_{ng} = \sqrt{N}\right)    
\end{align*}
}
The prior and hyperprior structure on $A$ is defined in Section \ref{sec:defRA}. From here on, let $\theta$ to refer to all prior and hyperprior parameters, and assume the parameter order for Truncated Normal distributions $\text{TruncNorm}(\mu, \sigma^2, a, b)$ where $a$ and $b$ are truncation bounds.

This model has $KN + NG$ parameters and $2(KN + NG)$ prior parameters to update on each iteration. MH-within-Gibbs steps are used for $P$ and $E$, meaning $KN + NG$ acceptance ratios must be computed.

\subsubsection{Prior Parameter Gibbs Updates: Full Conditional Distributions}

{\footnotesize
\begin{align*}
     \mu^P_{kn}|... &\sim \text{Normal}\left(\mu = \frac{m^P_{kn}/s^E_{kn} + P_{kn}/(\sigma^{P}_{kn})^2}{1/s^E_{kn} + 1/(\sigma^{P}_{kn})^2}, \sigma^2 = \frac{1}{{1/s^E_{kn} + 1/(\sigma^{P}_{kn})^2}}\right)\\
     (\sigma^{P}_{kn})^2|... &\sim \text{InverseGamma}\left(\alpha = a^P_{kn} + \frac{1}{2}, \beta = b^P_{kn} + \frac{1}{2}(P_{kn} - \mu^P_{kn})^2\right)\\
     \mu^E_{ng}|... &\sim \text{Normal}\left(\mu = \frac{m^E_{ng}/s^E_{ng} + E_{ng} / (\sigma^{E}_{ng})^2}{1/s^E_{ng} + 1/(\sigma^{E}_{ng})^2}, \sigma^2 = \frac{1}{{1/s^E_{ng} + 1/(\sigma^{E}_{ng})^2}}\right)\\
     (\sigma^{E}_{ng})^2 | ... &\sim \text{InverseGamma}\left(\alpha = a^E_{ng} + \frac{1}{2}, \beta = b^E_{ng} + \frac{1}{2}(E_{ng} - \mu^E_{ng})^2\right)
\end{align*}
}
The update for $R$ is defined in Section \ref{sec:updateRA}. 

\subsubsection{Parameter Gibbs Updates}
The update for $A_{nn}$ is defined in Section \ref{sec:updateRA}. 

\subsubsection{Parameter Gibbs Updates: MH-Within-Gibbs steps}

\underline{Targets}
{\footnotesize
\begin{align*}
    f(P^*_{kn}|P, E, M, \theta) &\propto  p_{\text{TruncNorm}}\left(P^*_{kn}|\mu^P_{kn}, (\sigma^P_{kn})^2, 0, \infty\right) \prod_g p_{\text{Poisson}}\left(M_{kg} | \lambda = (P^*E)_{kg}\right)\\
    f(E^*_{ng}|P, E, M, \theta) &\propto  p_{\text{TruncNorm}}\left(E^*_{ng}|\mu^E_{ng}, (\sigma^E_{ng})^2, 0, \infty\right) \prod_k p_{\text{Poisson}}\left(M_{kg} | \lambda = (PE^*)_{kg}\right)
\end{align*}
}

\noindent \underline{Proposals}
{\footnotesize
\begin{align*}
    g(P^*_{kn}|P, E, M, \theta) & \propto  p_{\text{TruncNorm}}(P^*_{kn}|\mu^P_{kn}, (\sigma^P_{kn})^2, 0, \infty) \prod_g p_{\text{Normal}}\left(M_{kg} | \mu = (P^*E)_{kg}, \sigma^2 = (PE)_{kg}\right)\\
    P^*_{kn}|P, E, M, \theta &\sim \text{TruncNorm}\left(\mu = m^P_{kn}(M,P,A,E,\theta), \sigma^2 = s^P_{kn}(M,P,E,\theta), 0, \infty\right)\\
    & m^P_{kn}(M,P,A,E,\theta) = \frac{\mu^P_{kn}/(\sigma^{P}_{kn})^2 + \sum_{g = 1}^G A_{nn}E_{ng}(M_{kg} - \hat M_{kg/n})/(PE)_{kg}}{1/(\sigma^{P}_{kn})^2 + \sum_{g = 1}^G A_{nn}E_{ng}^2/(PE)_{kg}}\\
    & s^P_{kn}(M,P,A,E,\theta) = \frac{1}{1/(\sigma^{P}_{kn})^2 + \sum_{g = 1}^G A_{nn}E_{ng}^2/(PE)_{kg}}\\\\
    g(E^*_{ng}|P, E, M, \theta) &\propto p_{\text{TruncNorm}}(E^*_{ng}|\mu^E_{ng}, (\sigma^E_{ng})^2, 0, \infty) \prod_k p_{\text{Normal}}\left(M_{kg} | \mu = (PE^*)_{kg}, \sigma^2 = (PE)_{kg}\right)\\
    E^*_{ng}|P, E, M, \theta&\sim \text{TruncNorm}\left(\mu = m^E_{ng}(M,P,A,E,\theta), \sigma^2 = s^E_{ng}(M,P,A,E,\theta), 0, \infty\right)\\
    &m^E_{ng}(M,P,A,E,\theta) = \frac{\mu^E_{ng}/(\sigma^{E}_{ng})^2 + \sum_{k = 1}^K P_{kn}A_{nn}(M_{kg} - \hat M_{kg/n})/(PE)_{kg}}{1/(\sigma^{E}_{ng})^2 + \sum_{k = 1}^K P_{kn}^2A_{nn}/(PE)_{kg}}\\
    &s^E_{ng}(M,P,A,E,\theta) = \frac{1}{1/(\sigma^{E}_{ng})^2 + \sum_{k = 1}^K P_{kn}^2A_{nn}/(PE)_{kg}}
\end{align*}
}

\noindent \underline{Acceptance Ratios}
{\footnotesize
\begin{align*}
    a^P_{kn} &= \frac{f(P^*_{kn}|P_{-kn}, E, M, \theta)g(P_{kn}|P^*, E, M, \theta)}{f(P_{kn}|P_{-kn}, E, M, \theta)g(P^*_{kn}|P, E, M, \theta)} \\
    &= \prod_g\frac{p_{\text{Poisson}}\left(M_{kg}|\lambda = (P^*E)_{kg}\right) \cdot p_{\text{Normal}}\left(M_{kg}|\mu = (PE)_{kg}, \sigma^2 = (P^*E)_{kg}\right)}{p_{\text{Poisson}}\left(M_{kg}|\lambda = (PE)_{kg}\right) \cdot p_{\text{Normal}}\left(M_{kg}|\mu = (P^*E)_{kg}, \sigma^2 = (PE)_{kg}\right)}\\\\
    a^E_{ng} &= \frac{f(E^*_{ng}|P, E_{-ng}, M, \theta)g(E_{ng}|P, E^*, M, \theta)}{f(E_{ng}|P, E_{-ng}, M, \theta)g(E^*_{ng}|P, E, M, \theta)} \\
    &= \prod_g\frac{p_{\text{Poisson}}\left(M_{kg}|\lambda = (PE^*)_{kg}\right) \cdot p_{\text{Normal}}\left(M_{kg}|\mu = (PE)_{kg}, \sigma^2 = (PE^*)_{kg}\right)}{p_{\text{Poisson}}\left(M_{kg}|\lambda = (PE)_{kg}\right) \cdot p_{\text{Normal}}\left(M_{kg}|\mu = (PE^*)_{kg}, \sigma^2 = (PE)_{kg}\right)}
\end{align*}
}

\noindent \underline{Updates / MH steps}
{\footnotesize
\begin{align*}
    P_{kn} = \begin{cases}
        P_{kn}^*\text{ with probability }a^P_{kn}\\
        P_{kn}\text{ with probability }1-a^P_{kn}
    \end{cases}\quad\quad
    E_{ng} = \begin{cases}
        E_{ng}^*\text{ with probability }a^E_{ng}\\
        E_{ng}\text{ with probability }1-a^E_{ng}
    \end{cases}
\end{align*}
}

\newpage
\subsection{Poisson - Exponential + MH}\label{sec:P_E_MH}

\subsubsection{Model Definition}

{\footnotesize
\begin{align*}
    \text{\underline{Likelihood}: }& M_{kg} \sim \text{Poisson}\left(\sum_{n}P_{kn}A_{nn}E_{ng}\right)\\
    \text{\underline{Priors}: }& P_{kn} \sim \text{Exponential}\left(\lambda^P_{kn}\right)\\
    & E_{ng} \sim \text{Exponential}\left(\lambda^E_{ng}\right)\\
    \text{\underline{Hyperpriors}: }& \lambda^P_{kn} \sim \text{Gamma}\left(a^P_{kn} = 10\sqrt{N}, b^P_{kn} = 10\sqrt{\bar M} \right)\\
    &\lambda^E_{ng} \sim \text{Gamma}\left(a^E_{ng} = 10\sqrt{N}, b^E_{ng} = 10\sqrt{\bar M} \right)\\
\end{align*}
}
The prior and hyperprior structure on $A$ is defined in Section \ref{sec:defRA}. From here on, let $\theta$ to refer to all prior and hyperprior parameters.

This model has $KN + NG$ parameters and $KN + NG$ prior parameters to update on each iteration. MH-within-Gibbs steps are used for $P$ and $E$, meaning $KN + NG$ acceptance ratios must be computed.

\subsubsection{Prior Parameter Full Conditional Distributions}

{\footnotesize
\begin{align*}
     \lambda^P_{kn}|... &\sim \text{Gamma}\left(a^P_{kn} + 1, b^P_{kn} + P_{kn}\right)\\
     \lambda^E_{ng}|... & \sim \text{Gamma}\left(a^E_{ng} + 1, b^E_{ng} + E_{ng}\right)
\end{align*}
}
The update for $R$ is defined in Section \ref{sec:updateRA}. 

\subsubsection{Parameter Gibbs Updates}
The update for $A_{nn}$ is defined in Section \ref{sec:updateRA}. 

\subsubsection{Parameter Gibbs Updates: MH-Within-Gibbs steps}

\underline{Targets}
{\footnotesize
\begin{align*}
    f(P^*_{kn}|P, E, M, \theta) &\propto  p_{\text{Exp}}\left(P^*_{kn}|\lambda^P_{kn}\right) \prod_g p_{\text{Poisson}}\left(M_{kg} | \lambda = (P^*E)_{kg}\right)\\
    f(E^*_{ng}|P, E, M, \theta) &\propto  p_{\text{Exp}}\left(E^*_{ng}|\lambda^E_{ng}\right) \prod_k p_{\text{Poisson}}\left(M_{kg} | \lambda = (PE^*)_{kg}\right)
\end{align*}
}

\noindent \underline{Proposals}
{\footnotesize
\begin{align*}
    g(P^*_{kn}|P, E, M, \theta) & \propto  p_{\text{Exp}}(P^*_{kn}|\lambda^P_{kn}) \prod_g p_{\text{Normal}}\left(M_{kg} | \mu = (P^*E)_{kg}, \sigma^2 = (PE)_{kg}\right)\\
    P^*_{kn}|P, E, M, \theta&\sim \text{TruncNorm}\left(\mu = m^P_{kn}(M,P,A,E,\theta), \sigma^2 = s^P_{kn}(M,P,E,\theta), 0, \infty\right)\\
    & m^P_{kn}(M,P,A,E,\theta) = \frac{\sum_{g = 1}^G A_{nn}E_{ng}(M_{kg} - \hat M_{kg/n})/(PE)_{kg} - \lambda^P_{kn}}{\sum_{g = 1}^G A_{nn}E_{ng}^2/(PE)_{kg}}\\
    & s^P_{kn}(M,P,A,E,\theta) = \frac{1}{\sum_{g = 1}^G A_{nn}E_{ng}^2/(PE)_{kg}}\\\\
    g(E^*_{ng}|P, E, M, \theta) &\propto p_{\text{Exp}}(E^*_{ng}|\lambda^E_{ng}) \prod_k p_{\text{Normal}}\left(M_{kg} | \mu = (PE^*)_{kg}, \sigma^2 = (PE)_{kg}\right)\\
    E^*_{ng}|P, E, M, \theta&\sim \text{TruncNorm}\left(\mu = m^E_{ng}(M,P,A,E,\theta), \sigma^2 = s^E_{ng}(M,P,A,E,\theta), 0, \infty\right)\\
    &m^E_{ng}(M,P,A,E,\theta) = \frac{\sum_{k = 1}^K P_{kn}A_{nn}(M_{kg} - \hat M_{kg/n})/(PE)_{kg} - \lambda^E_{ng}}{\sum_{k = 1}^K P_{kn}^2A_{nn}/(PE)_{kg}}\\
    &s^E_{ng}(M,P,A,E,\theta) = \frac{1}{\sum_{k = 1}^K P_{kn}^2A_{nn}/(PE)_{kg}}
\end{align*}
}

\noindent \underline{Acceptance Ratios}
{\footnotesize
\begin{align*}
    a^P_{kn} &= \frac{f(P^*_{kn}|P_{-kn}, E, M, \theta)g(P_{kn}|P^*, E, M, \theta)}{f(P_{kn}|P_{-kn}, E, M, \theta)g(P^*_{kn}|P, E, M, \theta)} \\
    &= \prod_g\frac{p_{\text{Poisson}}\left(M_{kg}|\lambda = (P^*E)_{kg}\right) \cdot p_{\text{Normal}}\left(M_{kg}|\mu = (PE)_{kg}, \sigma^2 = (P^*E)_{kg}\right)}{p_{\text{Poisson}}\left(M_{kg}|\lambda = (PE)_{kg}\right) \cdot p_{\text{Normal}}\left(M_{kg}|\mu = (P^*E)_{kg}, \sigma^2 = (PE)_{kg}\right)}\\\\
    a^E_{ng} &= \frac{f(E^*_{ng}|P, E_{-ng}, M, \theta)g(E_{ng}|P, E^*, M, \theta)}{f(E_{ng}|P, E_{-ng}, M, \theta)g(E^*_{ng}|P, E, M, \theta)} \\
    &= \prod_g\frac{p_{\text{Poisson}}\left(M_{kg}|\lambda = (PE^*)_{kg}\right) \cdot p_{\text{Normal}}\left(M_{kg}|\mu = (PE)_{kg}, \sigma^2 = (PE^*)_{kg}\right)}{p_{\text{Poisson}}\left(M_{kg}|\lambda = (PE)_{kg}\right) \cdot p_{\text{Normal}}\left(M_{kg}|\mu = (PE^*)_{kg}, \sigma^2 = (PE)_{kg}\right)}
\end{align*}
}

\noindent \underline{Updates / MH steps}
{\footnotesize
\begin{align*}
    P_{kn} = \begin{cases}
        P_{kn}^*\text{ with probability }a^P_{kn}\\
        P_{kn}\text{ with probability }1-a^P_{kn}
    \end{cases}\quad\quad
    E_{ng} = \begin{cases}
        E_{ng}^*\text{ with probability }a^E_{ng}\\
        E_{ng}\text{ with probability }1-a^E_{ng}
    \end{cases}
\end{align*}
}

\newpage
\subsection{Poisson - Gamma} \label{sec:P_G}

\subsubsection{Model Definition}

{\footnotesize
\begin{align*}
    \text{\underline{Likelihood}: }& M_{kg} = \sum_n Z_{kng}\\
    & Z_{kng} \sim \text{Poisson}\left(P_{kn}A_{nn}E_{ng}\right)\\
    \text{\underline{Priors}: }& P_{kn} \sim \text{Gamma}\left(\alpha^P_{kn}, \beta^P_{kn}\right)\\
    & E_{ng} \sim \text{Gamma}\left(\alpha^E_{ng}, \beta^E_{ng}\right)\\
    \text{\underline{Hyperpriors}: }& \beta^P_{kn} \sim \text{Gamma}(a^P_{kn} = 10\sqrt{N}, b^P_{kn} = 10)\\
    &\alpha^P_{kn} \sim \text{Gamma}(c^P_{kn} = 10\sqrt{\bar M}, d^P_{kn} = 10)\\
    & \beta^E_{ng} \sim \text{Gamma}(a^E_{ng} = 10\sqrt{N}, b^E_{ng} = 10)\\
    &\alpha^E_{ng} \sim \text{Gamma}(c^E_{ng} = 10\sqrt{\bar M}, d^E_{ng} = 10)
\end{align*}
}
The prior and hyperprior structure on $A$ is defined in Section \ref{sec:defRA}. From here on, let $\theta$ to refer to all prior and hyperprior parameters.

This model has $KN + NG + KNG$ parameters and $2(KN + NG)$ prior parameters to update on each iteration.

\subsubsection{Prior Parameter Gibbs Updates: Full Conditional Distributions}

{\footnotesize
\begin{align*}
    \beta^P_{kn} & \sim \text{Gamma}(a^P_{kn}  + \alpha^P_{kn}, b^P_{kn} + P_{kn})\\
    p(\alpha^P_{kn}|...) &\propto \frac{(\beta^P_{kn})^{\alpha^P_{kn}}}{\Gamma(\alpha^P_{kn})} P_{kn}^{(\alpha^P_{kn} - 1)}\left(\alpha^P_{kn}\right)^{c^P_{kn}-1}e^{-d^P_{kn}\alpha^P_{kn}}\\
    \beta^E_{ng} & \sim \text{Gamma}(a^E_{ng} + \alpha^E_{ng}, b^E_{ng} + E_{ng})\\
    p(\alpha^E_{ng}|...) & \propto \frac{(\beta^E_{ng})^{\alpha^E_{ng}}}{\Gamma(\alpha^E_{ng})} E_{ng}^{(\alpha^E_{ng}-1)}\left(\alpha^E_{ng}\right)^{c^E_{ng}-1}e^{-d^E_{ng} \alpha^EE_{ng}}
\end{align*}
}
$\alpha^P_{kn}$ and $\alpha^E_{ng}$ are sampled using Adaptive Rejection Metropolis Sampling (ARMS) as implemented by the {\tt armspp} R software package \citep{bertolacci2019armspp}. The update for $R$ is defined in Section \ref{sec:updateRA}.

\subsubsection{Parameter Gibbs Updates}
{\footnotesize
\begin{align*}
    P_{kn} |\dots&\sim \text{Gamma}\left(\alpha^P_{kn} + \sum_gZ_{kng}, \beta^P_{kn} + A_{nn}\sum_gE_{ng}\right)\\
    E_{ng} |\dots&\sim \text{Gamma}\left(\alpha^E_{ng} + \sum_kZ_{kng}, \beta^E_{ng} + A_{nn}\sum_kP_{kn}\right)\\
    \vec Z_{k\cdot g} |\dots&\sim \text{Multinomial}\left(P_{k1}A_{11}E_{1g}, \dots, P_{kN}A_{NN}E_{Ng}\right)
\end{align*}
}
The update for $A_{nn}$ is defined in Section \ref{sec:updateRA}.

\newpage
\subsection{Poisson - Exponential} \label{sec:P_E}

\subsubsection{Model Definition}

{\footnotesize
\begin{align*}
    \text{\underline{Likelihood}: }& M_{kg} = \sum_n Z_{kng}\\
    & Z_{kng} \sim \text{Poisson}\left(P_{kn}A_{nn}E_{ng}\right)\\
    \text{\underline{Priors}: }& P_{kn} \sim \text{Exponential}\left(\lambda^P_{kn}\right)\\
    & E_{ng} \sim \text{Exponential}\left(\lambda^E_{ng}\right)\\
    \text{\underline{Hyperpriors}: }& \lambda^P_{kn} \sim \text{Gamma}\left(a^P_{kn} = 10\sqrt{N}, b^P_{kn} = 10\sqrt{\bar M} \right)\\
    &\lambda^E_{ng} \sim \text{Gamma}\left(a^E_{ng} = 10\sqrt{N}, b^E_{ng} = 10\sqrt{\bar M} \right)\\
\end{align*}
}
The prior and hyperprior structure on $A$ is defined in Section \ref{sec:defRA}. From here on, let $\theta$ to refer to all prior and hyperprior parameters.

This model has $KN + NG + KNG$ parameters and $KN + NG$ prior parameters to update on each iteration.

\subsubsection{Prior Parameter Gibbs Updates: Full Conditional Distributions}

{\footnotesize
\begin{align*}
     \lambda^P_{kn}|... &\sim \text{Gamma}\left(a^P_{kn} + 1, b^P_{kn} + P_{kn}\right)\\
     \lambda^E_{ng}|... & \sim \text{Gamma}\left(a^E_{ng} + 1, b^E_{ng} + E_{ng}\right)
\end{align*}
}
The update for $R$ is defined in Section \ref{sec:updateRA}. 

\subsubsection{Parameter Gibbs Updates}
{\footnotesize
\begin{align*}
    P_{kn}|\dots &\sim \text{Gamma}\left(1 + \sum_gZ_{kng}, \lambda^P_{kn} + A_{nn}\sum_gE_{ng}\right)\\
    E_{ng}|\dots &\sim \text{Gamma}\left(1 + \sum_kZ_{kng}, \lambda^E_{ng} + A_{nn}\sum_kP_{kn}\right)\\
    \vec Z_{k\cdot g}|\dots &\sim \text{Multinomial}\left(P_{k1}A_{11}E_{1g}, \dots, P_{kN}A_{NN}E_{Ng}\right)
\end{align*}
}
The update for $A_{nn}$ is defined in Section \ref{sec:updateRA}.

\newpage
\subsection{Normal - TruncNormal (TN)} \label{sec:N_TN}

\subsubsection{Model Definition}

{\footnotesize
\begin{align*}
    \text{\underline{Likelihood}: }& M_{kg} \sim \text{Normal}\left(\sum_{n}P_{kn}A_{nn}E_{ng}, \sigma^2_g\right)\\
    \text{\underline{Priors}: }& P_{kn} \sim \text{TruncNorm}\left(\mu^P_{kn}, (\sigma^P_{kn})^2, 0, \infty\right)\\
    & E_{ng} \sim \text{TruncNorm}\left(\mu^E_{ng}, (\sigma^E_{ng})^2, 0, \infty\right)\\
    & \sigma^2_g \sim \text{InverseGamma}(\alpha_g = 3, \beta_g = 3)\\
    \text{\underline{Hyperpriors}: }& \mu^P_{kn} \sim \text{Normal}\left(m^P_{kn} = 0, (s^2_{kn})^2 = \sqrt{\bar M/N} \right)\\
    &(\sigma^P_{kn})^2 \sim \text{InverseGamma}\left(a^P_{kn} = N + 1, b^P_{kn} = \sqrt{N}\right)\\
    &\mu^E_{ng} \sim \text{Normal}\left(m^E_{ng} = 0, (s^E_{ng})^2 = \sqrt{\bar M/N} \right)\\
    &(\sigma^E_{ng})^2 \sim \text{InverseGamma}\left(a^E_{ng} = N + 1, b^E_{ng} = \sqrt{N}\right) 
\end{align*}
}
The prior and hyperprior structure on $A$ is defined in Section \ref{sec:defRA}. From here on, let $\theta$ to refer to all prior and hyperprior parameters, and assume the parameter order for Truncated Normal distributions $\text{TruncNorm}(\mu, \sigma^2, a, b)$ where $a$ and $b$ are truncation bounds.

This model has $KN + NG + G$ parameters and $2(KN + NG)$ prior parameters to update on each iteration.

\subsubsection{Prior Parameter Gibbs Updates: Full Conditional Distributions}

{\footnotesize
\begin{align*}
     \mu^P_{kn}|... &\sim \text{Normal}\left(\mu = \frac{m^P_{kn}/s^E_{kn} + P_{kn}/(\sigma^{P}_{kn})^2}{1/s^E_{kn} + 1/(\sigma^{P}_{kn})^2}, \sigma^2 = \frac{1}{{1/s^E_{kn} + 1/(\sigma^{P}_{kn})^2}}\right)\\
     (\sigma^{P}_{kn})^2|... &\sim \text{InverseGamma}\left(\alpha = a^P_{kn} + \frac{1}{2}, \beta = b^P_{kn} + \frac{1}{2}(P_{kn} - \mu^P_{kn})^2\right)\\
     \mu^E_{ng}|... &\sim \text{Normal}\left(\mu = \frac{m^E_{ng}/s^E_{ng} + E_{ng} / (\sigma^{E}_{ng})^2}{1/s^E_{ng} + 1/(\sigma^{E}_{ng})^2}, \sigma^2 = \frac{1}{{1/s^E_{ng} + 1/(\sigma^{E}_{ng})^2}}\right)\\
     (\sigma^{E}_{ng})^2 | ... &\sim \text{InverseGamma}\left(\alpha = a^E_{ng} + \frac{1}{2}, \beta = b^E_{ng} + \frac{1}{2}(E_{ng} - \mu^E_{ng})^2\right)
\end{align*}
}
The update for $R$ is defined in Section \ref{sec:updateRA}. 

\subsubsection{Parameter Gibbs Updates}

{\footnotesize
\begin{align*}
    P_{kn}| ... & \sim \text{TruncNorm}\left(\mu = m^P_{kn}(M,A,E,\theta), \sigma^2 = s^P_{kn}(M,A,E,\theta), 0, \infty\right)\\
    & m^P_{kn}(M,A,E,\theta) = \frac{\mu^P_{kn}/(\sigma^{P}_{kn})^2 + \sum_{g = 1}^G A_{nn}E_{ng}(M_{kg} - \hat M_{kg/n})/\sigma^2_g}{1/(\sigma^{P}_{kn})^2 + \sum_{g = 1}^G A_{nn}E_{ng}^2/\sigma^2_g}\\
    & s^P_{kn}(M,A,E,\theta) = \frac{1}{1/(\sigma^{P}_{kn})^2 + \sum_{g = 1}^G A_{nn}E_{ng}^2/\sigma^2_g}\\\\
    E_{ng}|... &\sim \text{TruncNorm}\left(\mu = m^E_{ng}(M,P,A,\theta), \sigma^2 = s^E_{ng}(M,P,A,\theta), 0, \infty\right)\\
    &m^E_{ng}(M,P,A,\theta) = \frac{\mu^E_{ng}/(\sigma^{E}_{ng})^2 + \sum_{k = 1}^K P_{kn}A_{nn}(M_{kg} - \hat M_{kg/n})/\sigma^2_g}{1/(\sigma^{E}_{ng})^2 + \sum_{k = 1}^K P_{kn}^2A_{nn}/\sigma^2_g}\\
    &s^E_{ng}(M,P,A,\theta) = \frac{1}{1/(\sigma^{E}_{ng})^2 + \sum_{k = 1}^K P_{kn}^2A_{nn}/\sigma^2_g}\\
     \sigma^2_g|... &\sim \text{InverseGamma}\left(\alpha_g + K/2, \beta_g + \sum_k(M_{kg} - \hat M_{kg})^2/2\right)
\end{align*}
}
The update for $A_{nn}$ is defined in Section \ref{sec:updateRA}. 

\subsection{Normal - Exponential}\label{sec:N_E}

\subsubsection{Model Definition}

{\footnotesize
\begin{align*}
    \text{\underline{Likelihood}: }& M_{kg} \sim \text{Normal}\left(\sum_{n}P_{kn}A_{nn}E_{ng}, \sigma^2_g\right)\\
    \text{\underline{Priors}: }& P_{kn} \sim \text{Exponential}\left(\lambda^P_{kn}\right)\\
    & E_{ng} \sim \text{Exponential}\left(\lambda^E_{ng}\right)\\
    &\sigma^2_g \sim \text{InverseGamma}\left(\alpha_g = 3, \beta_g = 3\right)\\
    \text{\underline{Hyperpriors}: }& \lambda^P_{kn} \sim \text{Gamma}\left(a^P_{kn} = 10\sqrt{N}, b^P_{kn} = 10\sqrt{\bar M} \right)\\
    &\lambda^E_{ng} \sim \text{Gamma}\left(a^E_{ng} = 10\sqrt{N}, b^E_{ng} = 10\sqrt{\bar M} \right)
\end{align*}
}
The prior and hyperprior structure on $A$ is defined in Section \ref{sec:defRA}. From here on, let $\theta$ to refer to all prior and hyperprior parameters.

This model has $KN + NG + G$ parameters and $KN + NG$ prior parameters to update on each iteration.

\subsubsection{Prior Parameter Full Conditional Distributions}

{\footnotesize
\begin{align*}
     \lambda^P_{kn}|... &\sim \text{Gamma}\left(a^P_{kn} + 1, b^P_{kn} + P_{kn}\right)\\
     \lambda^E_{ng}|... & \sim \text{Gamma}\left(a^E_{ng} + 1, b^E_{ng} + E_{ng}\right)
\end{align*}
}
The update for $R$ is defined in Section \ref{sec:updateRA}. 

\subsubsection{Parameter Gibbs Updates}

{\footnotesize
\begin{align*}
    P_{kn}... & \sim \text{TruncNorm}\left(\mu = m^P_{kn}(M,A,E,\theta), \sigma^2 = s^P_{kn}(M,A,E,\theta), 0, \infty\right)\\
    & m^P_{kn}(M,A,E,\theta) = \frac{\sum_{g = 1}^G A_{nn}E_{ng}(M_{kg} - \hat M_{kg/n})/\sigma^2_g - \lambda^P_{kn}}{\sum_{g = 1}^G A_{nn}E_{ng}^2/\sigma^2_g}\\
    & s^P_{kn}(M,A,E,\theta) = \frac{1}{\sum_{g = 1}^G A_{nn}E_{ng}^2/\sigma^2_g}\\\\
    E_{ng}|... &\sim \text{TruncNorm}\left(\mu = m^E_{ng}(M,P,A,\theta), \sigma^2 = s^E_{ng}(M,P,A,\theta), 0, \infty\right)\\
    &m^E_{ng}(M,P,A,\theta) = \frac{\sum_{k = 1}^K P_{kn}A_{nn}(M_{kg} - \hat M_{kg/n})/\sigma^2_g - \lambda^E_{ng}}{\sum_{k = 1}^K P_{kn}^2A_{nn}/\sigma^2_g}\\
    &s^E_{ng}(M,P,A,\theta) = \frac{1}{\sum_{k = 1}^K P_{kn}^2A_{nn}/\sigma^2_g}\\
    \sigma^2_g|... &\sim \text{InverseGamma}\left(\alpha_g + K/2, \beta_g + \sum_k(M_{kg} - \hat M_{kg})^2/2\right)
\end{align*}
}
The update for $A_{nn}$ is defined in Section \ref{sec:updateRA}. 

\newpage
\section{Appendix B: Implementation Details}
\label{supp_sec:appendix_b}
\subsection{Inference}\label{sec:suppB_inference}

Inference is performed on the last 1000 posterior samples. No tempered or non-MH samples are used for final posterior inference. This same strategy is on periodic windows of samples to check for convergence.

When learning rank, we first compute $\hat A$, the maximum a-posteriori (MAP) estimate of the signature inclusion matrix, at its mode. We then subset inference samples to those that match this MAP.

To summarize the inference samples, we next remove the scale indeterminacy of NMF by normalizing each iteration's $P^{(i)}$ and $E^{(i)}$ matrices for columns of $P^{(i)}$ to sum to one:
\begin{align*}
    E'_{ng} = E_{ng} * \sum_kP_{kn}\\
    P'_{kn} = P_{kn} / \sum_{k''}P_{k''n}.
\end{align*}
This rescaling ensures the reconstruction has not changed, $PE = P'E'$.

Then, final estimates $\hat P$ and $\hat E$ and 95\% credible intervals are element-wise means and 2.5th and 97.5th percentiles, respectively. The estimated reconstruction $\hat M$ is the product $\hat P \hat A \hat E$.

\subsection{Convergence}\label{sec:convergence}

Common methods to check convergence of MCMC methods pose three challenges in the context of latent factor models: first, there are thousands of parameters of interest (the number of elements of the $P$ and $E$ matrices), which means an infeasible number of traceplots to visually inspect. Second, the scale non-identifiability of NMF (e.g., $\hat{P} \hat{E} = (2\hat{P}) (\frac{1}{2}\hat{E})$) means that trace plots of values in $P$ and $E$ may have a slope, and thus not seem to have converged, even if the product $PE$ is a stable solution. Third, learned latent factors may not be the same across multiple chains, so methods that compare across-chain to within-chain variation cannot be used. 

We instead define a programmatic convergence criterion that is appropriate for any scale of data and ensures that the various models reach the same level of convergence. This convergence criterion is defined on a running-metrics plot, built off of the concept of a running means plot. We assume that if posterior samples have converged, then the average over windows of fixed lengths should not change as the window shifts, and thus a scalar metric of this average should not change either. 

Each time convergence is checked, inference over the window follows Section \ref{sec:suppB_inference}. Metrics RMSE and KL-Divergence are computed on the reconstruction $\hat M = \hat P \hat A \hat E$. However, scale-dependent metrics of log-likelihood, BIC, and log-posterior don't make sense to compute on the rescaled $\hat P$ and $\hat E$. Instead, these window metrics are the average over per-sample metrics within that window.

All parameters can be adjusted as input to the {\tt new\_convergence\_controls} function. Default settings used for all results in our paper consider the convergence of log posterior ({\tt metric}) with a window size of $1000$ samples ({\tt MAP\_over}) and window shifts of $100$ samples ({\tt MAP\_every}). The package also allows for convergence of BIC, log likelihood, RMSE, or KL-divergence. A change of less than $0.1$\% in log posterior is considered ``no change'' ({\tt tol}). We only check for convergence once the sampler has run for at least a minimum number of iterations ({\tt miniters}) and none of the samples in the window were tempered. We say a sampler has converged if the log posterior has not changed in the past 5 status checks ({\tt Ninarow\_nochange}), has not improved in the past 10 status checks ({\tt Ninarow\_nobest}), or has reached the maximum number of iterations specified ({\tt maxiters}).

\begin{figure}
    \centering
    \includegraphics[width=0.7\linewidth]{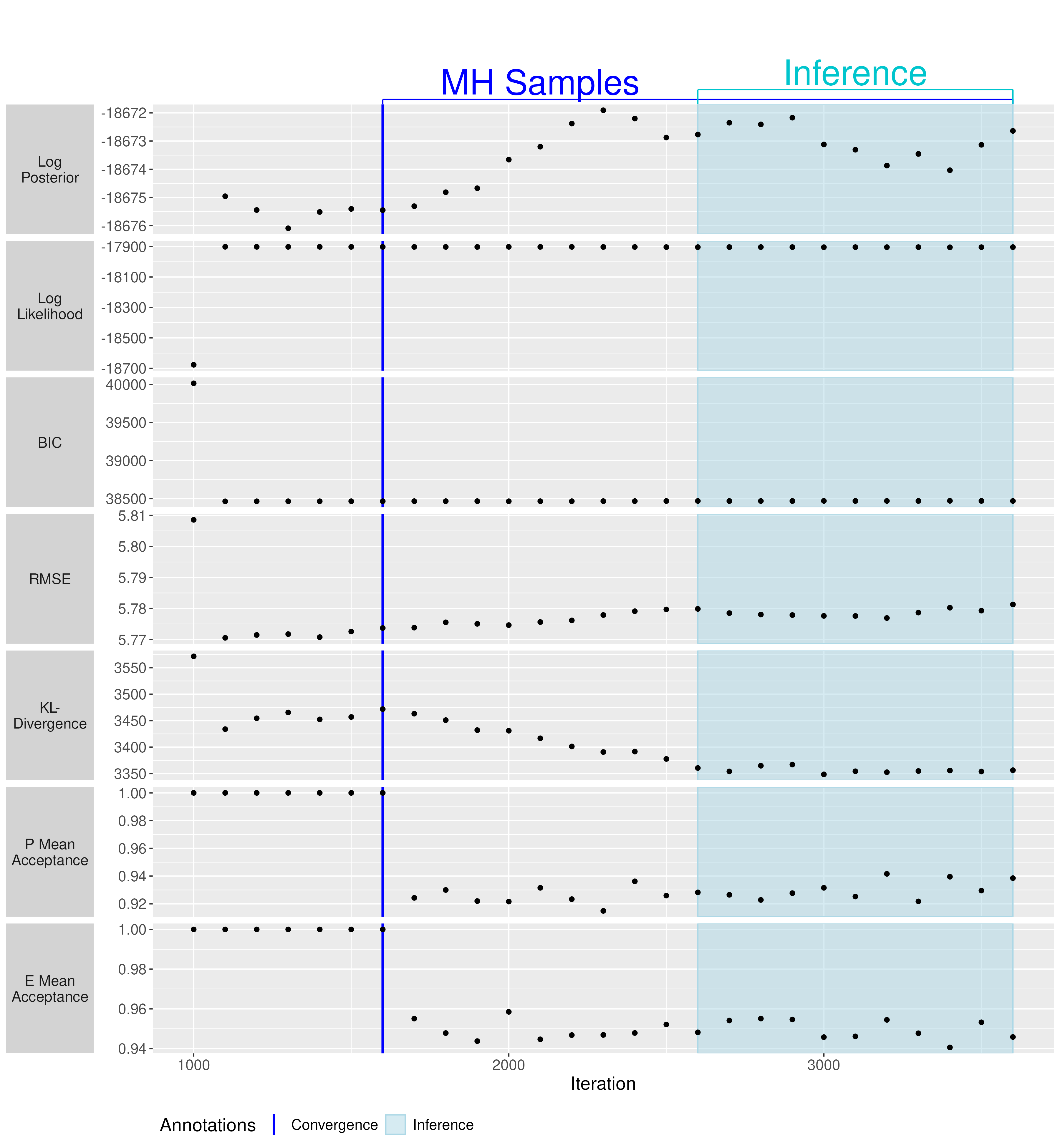}
    \caption{Running metrics plot for an example simulated dataset with fixed rank. Vertical blue line indicates convergence (see Section \ref{sec:convergence}) and thus the switch from ``accept all'' to true MH samples (see Section \ref{sec:warmup}). The highlighted light blue rectangle indicates samples used for final inference.}
    \label{fig:trace_plot_MAP}
\end{figure}

Figure \ref{fig:trace_plot_MAP} shows a running metrics plot using the default convergence control parameters on an example simulated dataset with fixed rank. Note the extremely tight y-axis for log-posterior distribution (was NA at iteration 1000). 

\subsection{Warmup and MH Steps}\label{sec:warmup}

We find in practice that the MH-within-Gibbs steps are slow to explore when starting from prior samples. Instead, we accept all proposals in a ``warm-up'' phase to find the high-posterior space shared by the Normal and Poisson Bayesian NMFs. Once the sampler has converged, it switches to true MH steps for the last 2000 iterations, using only the last 1000 for inference.

We see in practice that after switching to true MH steps, the RMSE increases slightly and KL-Divergence decreases slightly (Figure \ref{fig:trace_plot_MAP}). This makes sense as we move away from samples closer to a Normal Bayesian NMF posterior space (whose likelihood optimizes RMSE) to a Poisson Bayesian NMF posterior space (whose likelihood optimizes KL-Divergence).

\subsection{Alignment to Reference Signatures} \label{sec:alignment}

In order to interpret and evaluate results, we frequently wish to compare estimated factors (columns of $\hat P$) to reference factors (columns of $P$) from previous analysis or domain knowledge, commonly from the COSMIC database \citep{icgc2020pan} in the case of mutational signatures. We use the following approach to align estimated signatures to ground truth in our simulation studies as well.

We follow existing mutational signatures works \citep{rosales2016, gori2018sigfit, alexandrov2020repertoire, grabski2025bayesian} and focus on the scale-independent measure of cosine similarities between signatures. To align estimated factors to the reference factors, we first create a similarity matrix $S$, where each element is the cosine similarity between a column of $P$ and a column of $\hat P$, $S_{ij} = cos(P_i, \hat{P}_j)$. The Hungarian algorithm \citep{kuhn1955hungarian} is used on $-1\cdot S$, which solves the combinatorial optimization problem to reorder rows and columns to minimize the sum of values on the diagonal. The resulting aligned similarity matrix, $S^*$, has an optimal alignment in terms of total aligned cosine similarity. This approach only allows a single estimated signature to be aligned to each reference signature, and vice versa.

We incorporate posterior uncertainty into this process by performing factor alignment for each of the posterior samples used to compute $\hat P$. We determine final alignments through majority voting with cosine similarity as voting weights.

\subsection{Label Switching Diagnostic}

Label switching is always a concern in Bayesian mixture models. This is partially mitigated by our use of hyperpriors, which avoids factors fully resetting to the prior when they are excluded. However, even with hyperpriors, label switching is still possible. To account for this, we have added additional visualization capabilities to our R software package to help diagnose label switching. 

Given a set of reference factors (e.g., COSMIC mutational signatures), we visualize the closest match of each signature across iterations. If label switching occurs, the color of a latent factor will switch part way through the chain. 

\begin{figure}
    \centering
    \includegraphics[width = \linewidth]{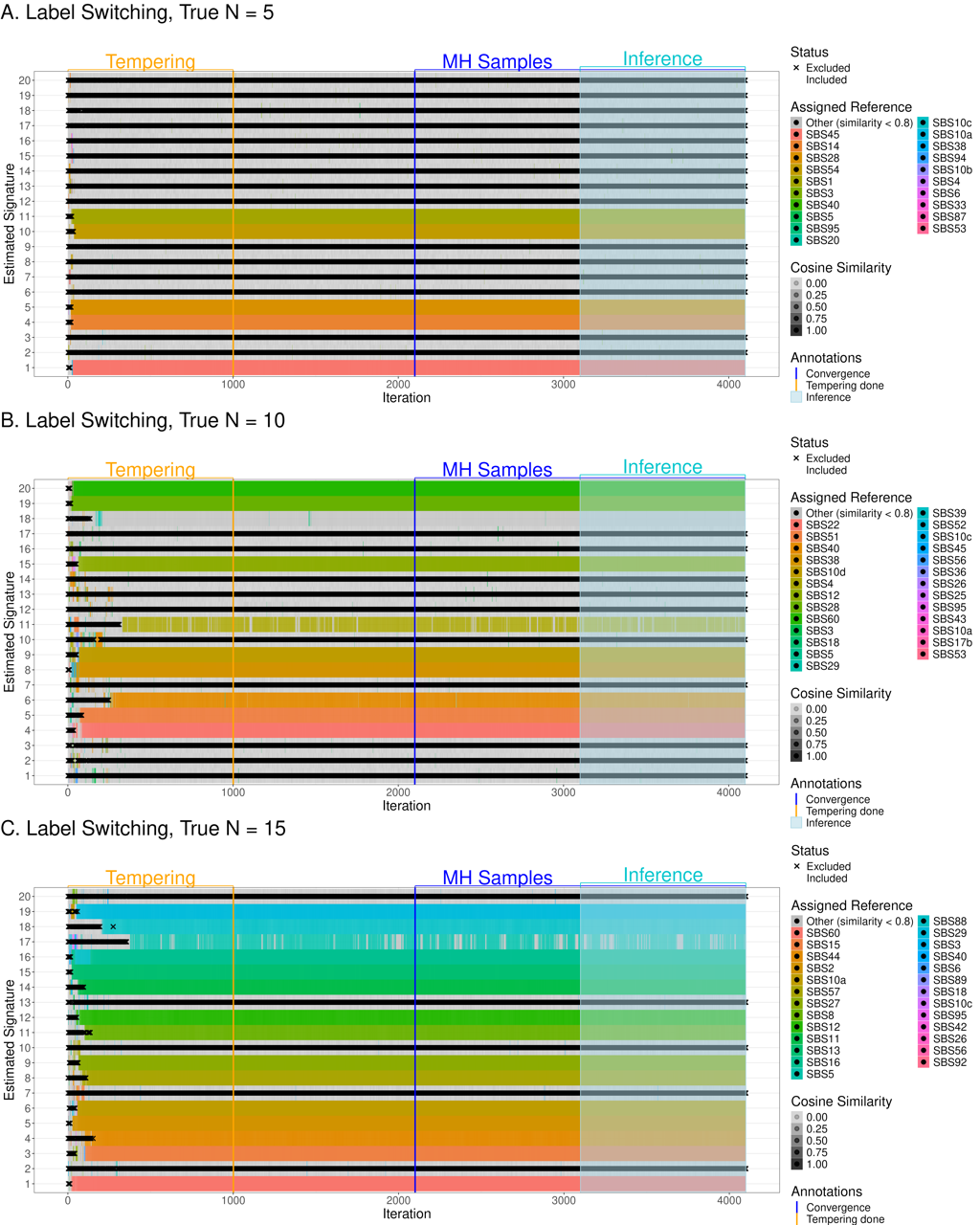}
    \caption{Label switching diagnostic plot on simulated data with \textbf{A.} N = 5, \textbf{B.} N = 10, and \textbf{C.} N = 15 signatures.}
    \label{fig:label_switching}
\end{figure}

In our simulation studies, we see evidence of label switching only very early on in the warm-up phase (first 500 iterations or less), but labels stabilize far before samples are used for inference. For example, in Figure \ref{fig:label_switching}B, There is some label switching with SBS40, which is initially aligned to estimated signature 14, then estimated signature 10, and eventually estimated signature 6 within the first few hundred iterations. SBS40 is then consistently assigned to estimated signature 6 for the remainder of the sampler's run.

Although these individual observations do not constitute a full benchmarking, this functionality provides a diagnostic for individual use cases.

\newpage
\section{Appendix C: Additional Simulation Results}
\label{supp_sec:appendix_c}

\subsection{Fixed Rank Simulation Additional Results}\label{sec:supp_study1_metrics}

\begin{figure}
    \centering
    \includegraphics[width=0.8\linewidth]{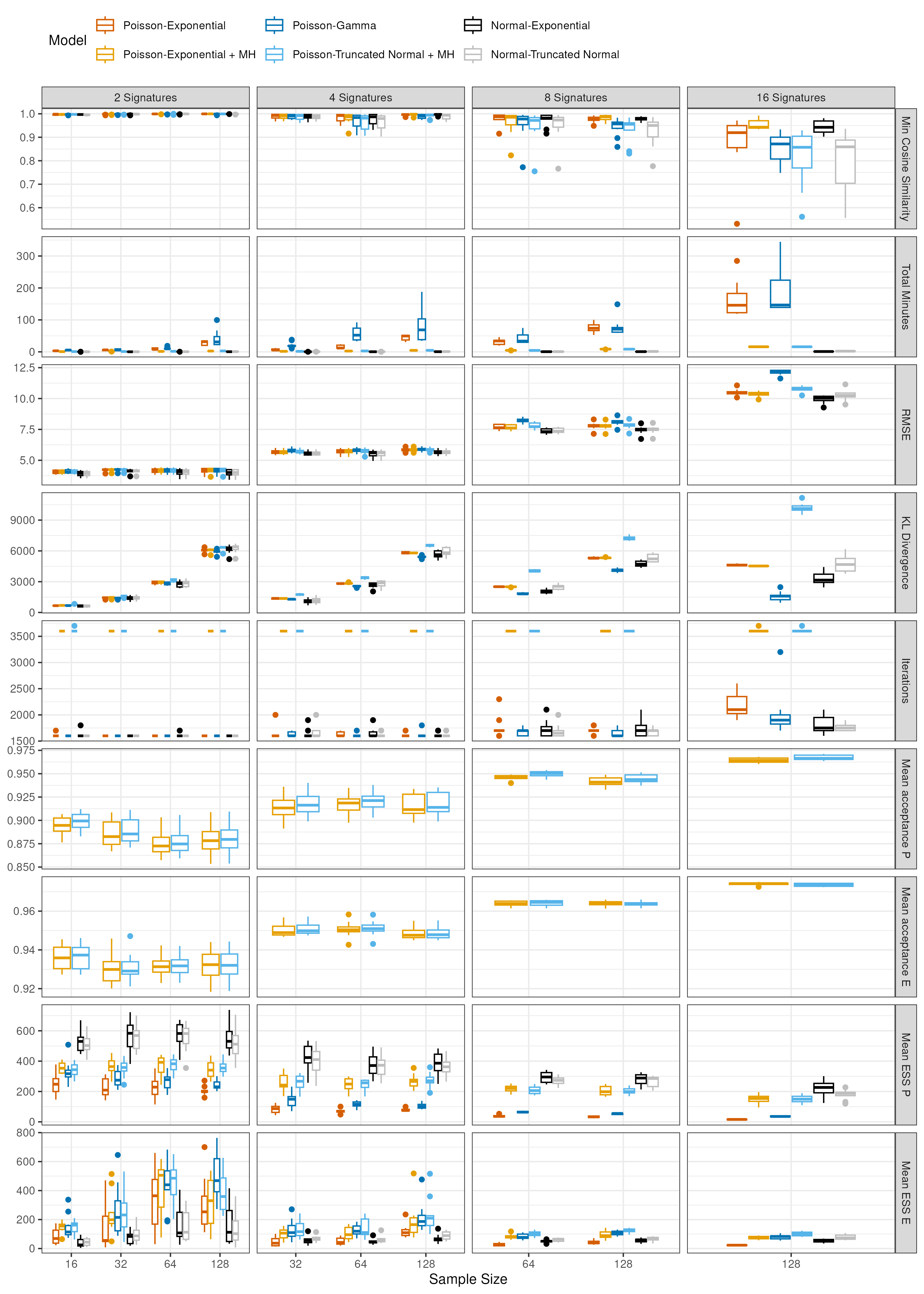}
    \caption{Metrics for all models on simulated data with fixed rank, including average acceptance rate across all elements of the P and E matrices for Poisson+MH models, and average effective sample size (ESS) across all elements of the P and E matrics (out of 1000 posterior samples).}
    \label{fig:metrics}
\end{figure}

Figure \ref{fig:metrics} is an extended version of main text Figure 2, reporting metrics for all six models. Warm colors are Poisson-Exponential models with and without MH steps. We expect these to be near identical in minimum cosine similarity, RMSE, and KL-Divergence. Cool colors are Poisson-Gamma (with augmentation) and Poisson-Truncated Normal (with MH steps). We expect these to be close, but not identical because of the different priors used. In black and grey are Normal-likelihood models.

In addition to metrics discussed in the main text, we report mean acceptance rates and effective sample sizes. Mean acceptance rates are only reported for Poisson+MH models, and report the average acceptance ratio over all elements of the P or E matrix, averaged again over all 1000 posterior samples. The results show that acceptance rates are very high ($\ge 0.9$, on average), consistent with expectations for high-overlap, geometry-informed proposals.

Effective sample size (ESS), or the approximate number of independent samples that would correspond to a dependent chain of values $\mathbf x$, is estimated as follows:
\begin{align*}
    \widehat{ESS}(\mathbf x) &= \frac{n}{1 + 2 \sum_{k = 1}^K\rho(k)}\\
    \rho(k) & = \text{cor}\left(\mathbf x_{1:(n-k)}, \mathbf x_{k:n}\right)\\
    K &= \arg \min_k \{k: \rho(k) < 0\}
\end{align*}
where $n$ is the length of vector $\mathbf x$, $\rho(k)$ is the correlation between $\mathbf x$ and itself at a lag of $k$, and $K$ is the smallest value of $k \in 1,..., n$ that results in a negative lag correlation $\rho(k)$.

Effective sample sizes tends to decrease with dimensionality. This is expected, as conditional dependencies tighten with more parameters (see \citet{betancourt2017conceptual} for an intuitive explanation). We also see a pattern of higher ESS for our Poisson+MH models than their corresponding standard Poisson samplers, indicating a more efficient exploration of the posterior space.

\subsection{Comparing Point Estimates: Additional Details}

\begin{table}[]
    \centering
    \begin{tabular}{|c|c|c|c|}
        \hline
        Comparison & Prior & \# datasets with & \# datasets with \\
         &  & P min cosine sim $<$0.8 & E min cosine sim $<$0.8\\
        \hline
        Normal vs Poisson&Exponential&5&2\\
        Normal vs Poisson&Gamma / TN&5&1\\
        Poisson vs Poisson+MH&Exponential&6&2\\
        Poisson vs Poisson+MH&Gamma / TN&3&0\\
        Poisson vs Poisson&Exponential&7&3\\
        Poisson vs Poisson&Gamma / TN&1&1\\
        \hline
    \end{tabular}
    \caption{Number of points excluded from Figure 1A by comparison and prior, out of 100 datasets.}
    \label{tab:excluded_fig1a}
\end{table}

Table \ref{tab:excluded_fig1a} shows the number of points excluded from main text Figure 1A. Importantly, the number is comparable across all comparisons, even the null case comparing two chains of the standard Poisson sampler. This indicates that such deviation is reasonably expected due to the randomness of Gibbs samplers, and does not change the results reported in the main text.

\subsection{Comparing Posterior Uncertainty}

\begin{figure}
    \centering
    \includegraphics[width=\linewidth]{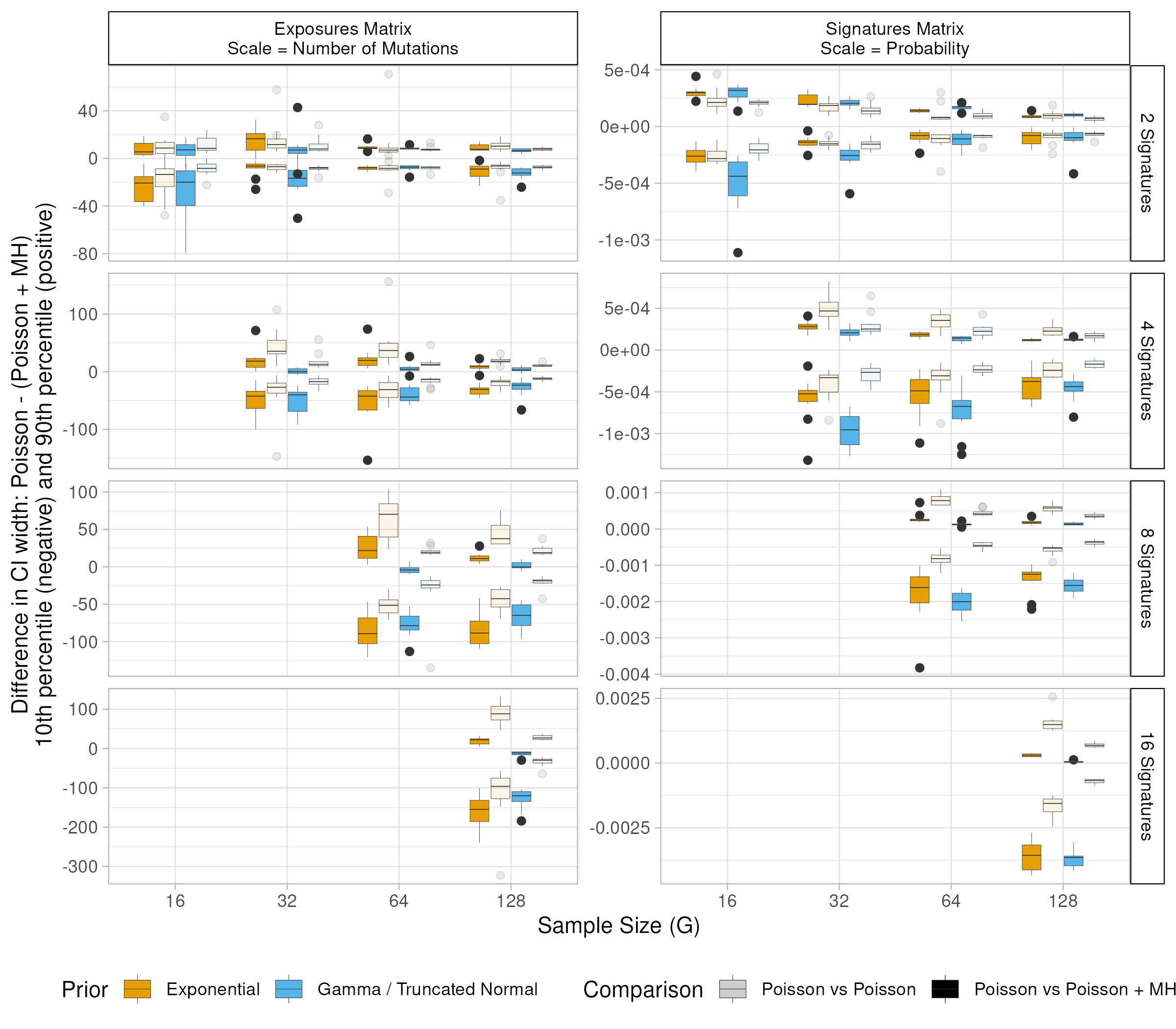}
    \caption{Posterior uncertainty agreement between Poisson Bayesian NMF models in exposures matrix (left) and signatures matrix (right). Plotted are 10th (negative) and 90th (positive) percentiles of the difference in credible interval widths with standard Poisson as reference. Transparent boxes represent the null case comparing two chains of a standard Poisson sampler. Colored boxes compare Poisson + MH to standard Poisson.}
    \label{fig:credible_intervals}
\end{figure}

\label{sec:comp_post}
\subsubsection{Methods}

Each sampler S returns point estimates as well as 95\% credible interval bounds for elements of P, $[l^S(P_{kn}), u^S(P_{kn})]$, and E, $[l^S(E_{ng}), u^S(E_{ng})]$.

To compare two samplers S and Q on a given simulated dataset, we first compute credible interval widths $w$ for each element of $P$ and $E$ from each sampler. For example, for element $P_{kn}$ ,:
\begin{align}
    w^S(P_{kn}) & = u^S(P_{kn}) - l^S(P_{kn}), \quad w^Q(P_{kn}) = u^Q(P_{kn}) - l^Q(P_{kn}).
\end{align}
Next, we compute element-wise differences in widths between samplers, $d$:
\begin{align}
    d^{SQ}(P_{kn})& = w^S(P_{kn}) - w^Q(P_{kn}).
\end{align}

Letting $d^{SQ}(P)$ be the flattened vector of $d^{SQ}(P_{kn})$ for all $k,n$, and similarly for $d^{SQ}(E)$, we compute the 10th and 90th percentiles of these differences, separately for $P$ and $E$. Then, comparing samplers S and Q on the given dataset can be summarized by two ranges:
\begin{align}
    [\text{perc}_{10}(d^{SQ}(P)), \text{perc}_{90}(d^{SQ}(P))]\\
    [\text{perc}_{10}(d^{SQ}(E)), \text{perc}_{90}(d^{SQ}(E))]
\end{align}

Treating sampler S as a reference, these can be interpreted as reasonable ranges of how conservative (wider, lower limits) or anticonservative (narrower, upper limits) sampler Q is expected to be compared to sampler S. Narrow ranges around zero indicate a similar degree of posterior uncertainty between samplers---this, combined with near-identical point point estimates, indicates the two samplers explore the same posterior space.

\subsubsection{Simulation Studies}

In our simulation studies with known rank (see main body Section 5.3), we compare posterior uncertainties as described above between the Poisson + MH sampler and the standard Poisson sampler. We also compare two chains of the standard Poisson sampler to describe a null case, or the range of differences expected due to the randomness of samplers.

For each comparison $S, Q$, Figure \ref{fig:credible_intervals} plots the ranges $[\text{perc}_{10}(d^{SQ}(P)), \text{perc}_{90}(d^{SQ}(P))]$ (left) and $[\text{perc}_{10}(d^{SQ}(E)), \text{perc}_{90}(d^{SQ}(E))]$ (right) across the 100 datasets. Transparent boxes represent the null case. Instances where the colored boxes fall below the transparent boxes indicate that our Poisson + MH model is more conservative than expected. 

We see that Poisson + MH samplers are generally a bit more conservative than standard Poisson samplers, and are only slightly anticonservative in the simplest cases when latent rank is 2. We also see that when the prior is an exact match (Exponential, orange), the widths are much closer to the expected variation. This is reasonable because the posterior space \textit{should} change when the prior changes from a Gamma to a Truncated Normal (blue). 

Overall, the scale of the y-axes indicate differences are minute, with all signature widths within 0.004 (on the probability scale) and all exposures within 2\% of total mutations per sample. The difference in mutations ranges from 20 mutations per sample (seen for individuals with 2 signatures and thus an expected 2,000 mutations) to 200 mutations per sample (seen for individuals with 16 signatures and thus an expected 16,000 mutations).

The slightly wider credible intervals may in fact be a result of more efficient posterior exploration as seen with higher effective sample sizes for Poisson+MH in Section \ref{sec:supp_study1_metrics}. This suggests these wider credible intervals may be more accurate, while standard Poisson samplers may be too narrow.

\subsection{Learning Rank Simulation Additional Results}

\begin{figure}
    \centering
    \includegraphics[width = \linewidth]{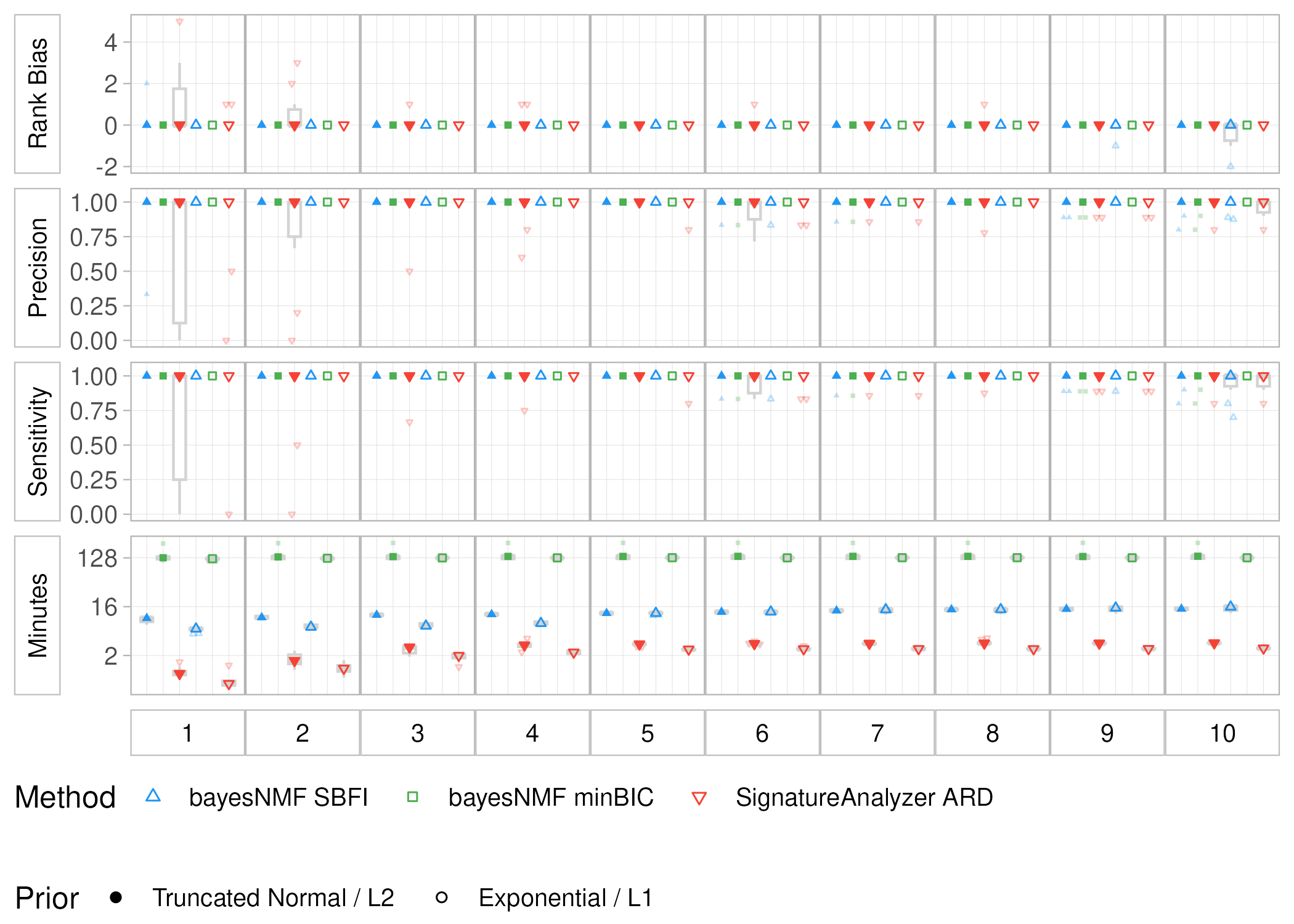}
    \includegraphics[width = \linewidth]{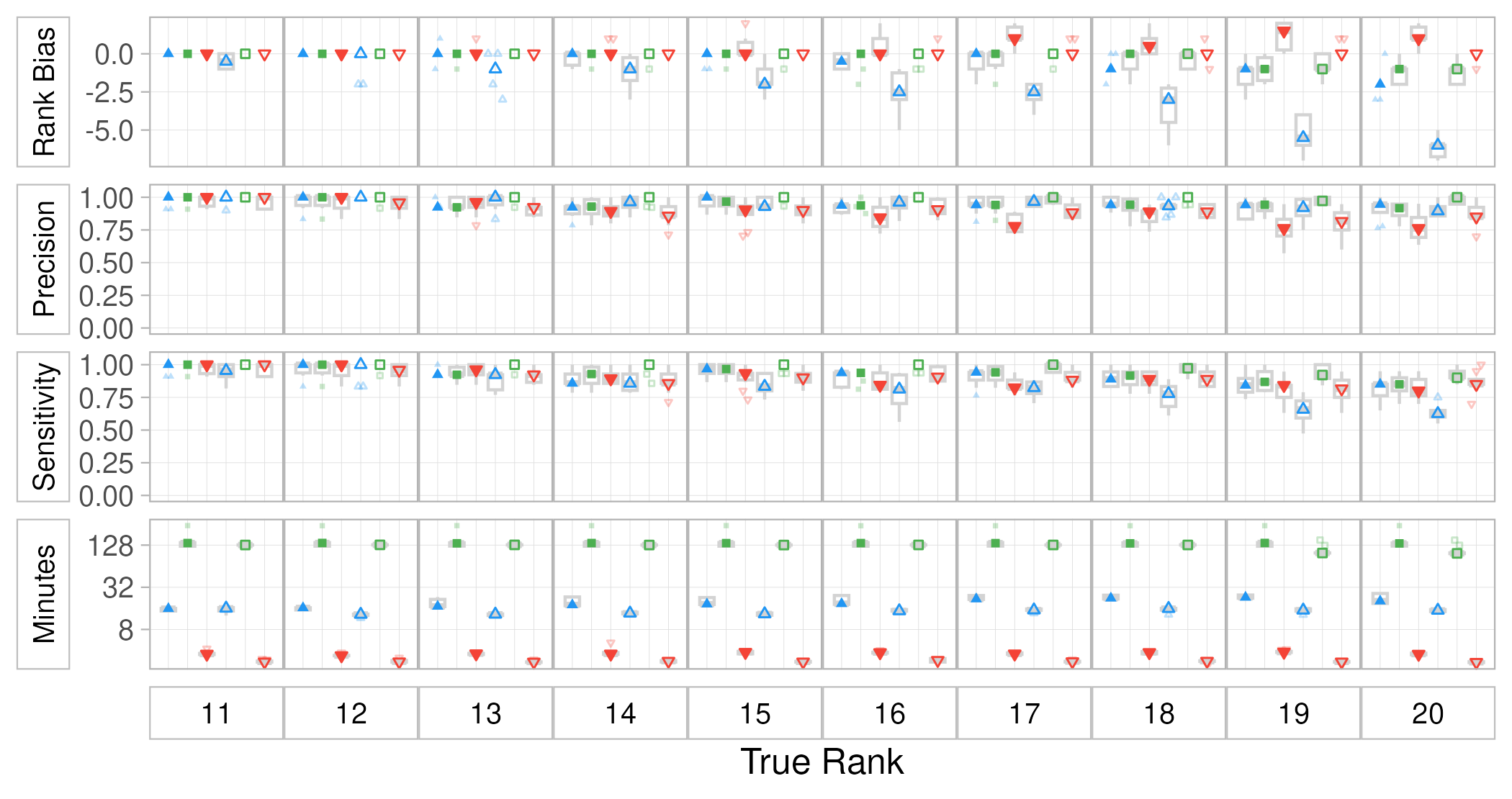}
    \caption{Rank bias, precision, sensitivity, and time comparison of rank learning approaches: {\tt bayesNMF} with SBFI (upright triangle), and Heuristic (square), as well as {\tt SignatureAnalyzer}+ARD (inverted triangle). Plotted are large dark points for medians overlapping lighter boxplots and jittered outliers. Split into N = 1-10 (top) and 11-20 (bottom).}
    \label{fig:learn-rank-res-extra}
\end{figure}

Figure \ref{fig:learn-rank-res-extra} is an extended version of main text Figure 3, reporting metrics for all ranks between 1 and 20. Results match what is reported in main text 5.3.

\subsection{Sensitivity Analysis: Sparse Data}

Because the Normal approximation to Poisson breaks down with sparse or low counts, we investigate how bayesNMF performs on this type of data.

\subsubsection{Fixed Rank}\label{sec:sparse_fixed}

\begin{figure}
    \centering
    \includegraphics[width = \linewidth]{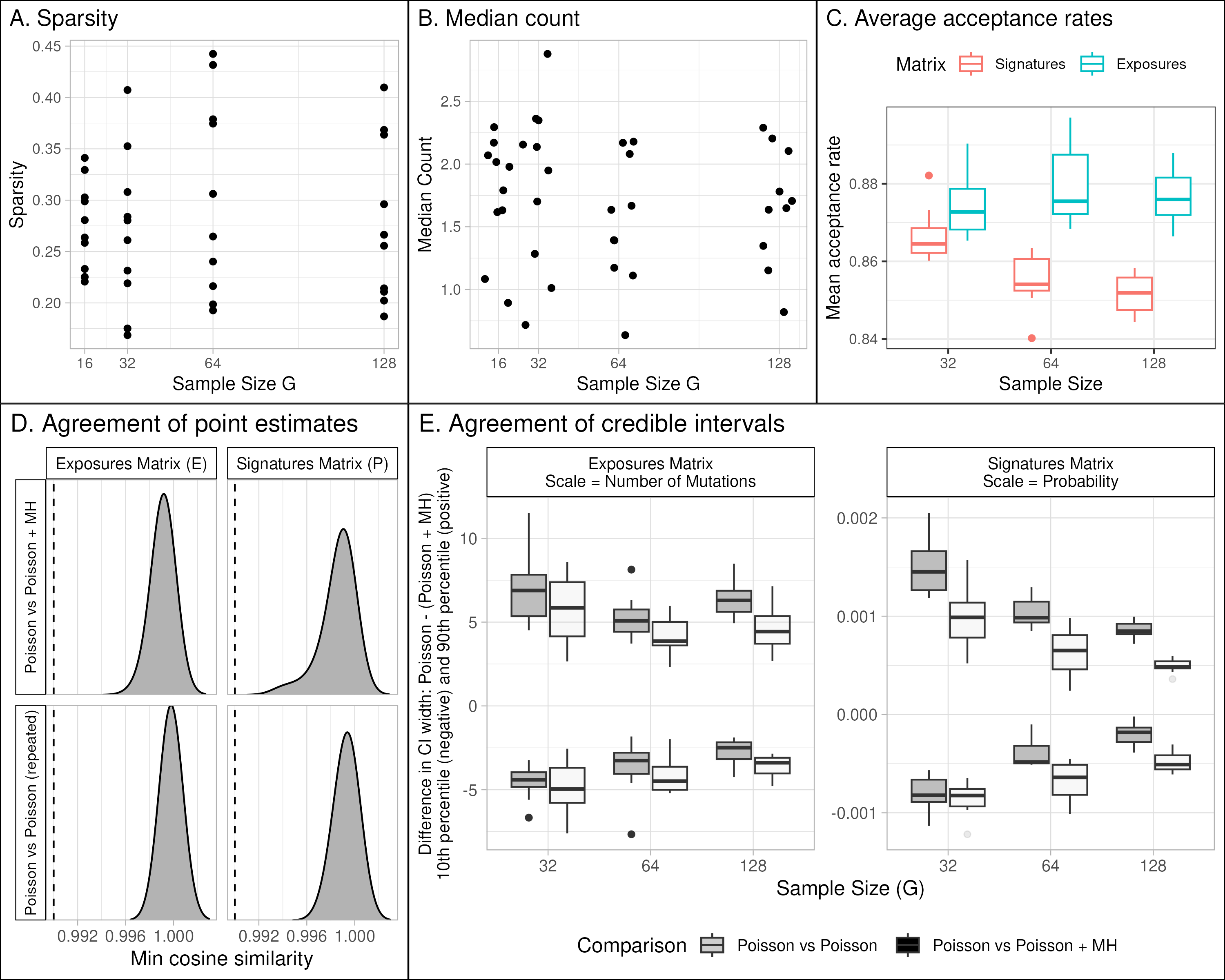}
    \caption{Data composition and results of bayesNMF with Exponential priors using fixed rank. \textbf{A.} Sparsity (proportion 0s) of simulated data. \textbf{B.} Median of mutational counts matrix of simulated datasets. \textbf{C.} Mean acceptance rates within each matrix across posterior samples used for inference. \textbf{D.} Agreement of point estimates between Poisson and Poisson + MH and the null case between repeated Poisson chains. \textbf{E.} Agreement of 95\% credible intervals in terms of 10 and 90th percentile of width difference (see Section \ref{sec:comp_post}).}
    \label{fig:study1_sparse}
\end{figure}

Data are simulated as in main text Section 5.2 for N = 4 signatures only and a reduced expected number of mutations per sample per signature of 100. This implies 400 mutations per sample spread across 96 mutation types, resulting in 20-40\% sparsity (Figure \ref{fig:study1_sparse}A) and low median mutational counts below 3 (Figure \ref{fig:study1_sparse}B). This analysis only compares Poisson-Exponential and Poisson-Exponential + MH (focusing on the case of exact model alignment). 

The average acceptance rate has decreased from around 0.9-0.95 in the primary simulation study to 0.85-0.9 in this sensitivity analysis (Figure \ref{fig:study1_sparse}C). This is expected because the geometry of the Normal NMF is not as close to the Poisson NMF when counts are low. However, these acceptance rates are still relatively high, indicating our proposals are still high-overlap and geometry-informed.

With these results, we recreated main text Figure 2 comparing point estimates, which shows that even with sparse counts, Poisson+MH reaches the same point estimate as a standard Poisson sampler (Figure \ref{fig:study1_sparse}D). We've also recreated Appendix Figure \ref{fig:credible_intervals}, which shows that the Poisson+MH reaches the same posterior uncertainty as a standard Poisson sampler (Figure \ref{fig:study1_sparse}E).

Together, these results show that even with sparse data, the Poisson-Exponential+MH model is still able to match the standard Poisson-Exponential model, both in terms of point estimates and posterior uncertainty.

\subsubsection{Learning Rank}

This sensitivity analysis compares Poisson+MH SBFI to SignatureAnalyzer with Truncated Normal/L2 priors. 

\begin{figure}
    \centering
    \includegraphics[width=\linewidth]{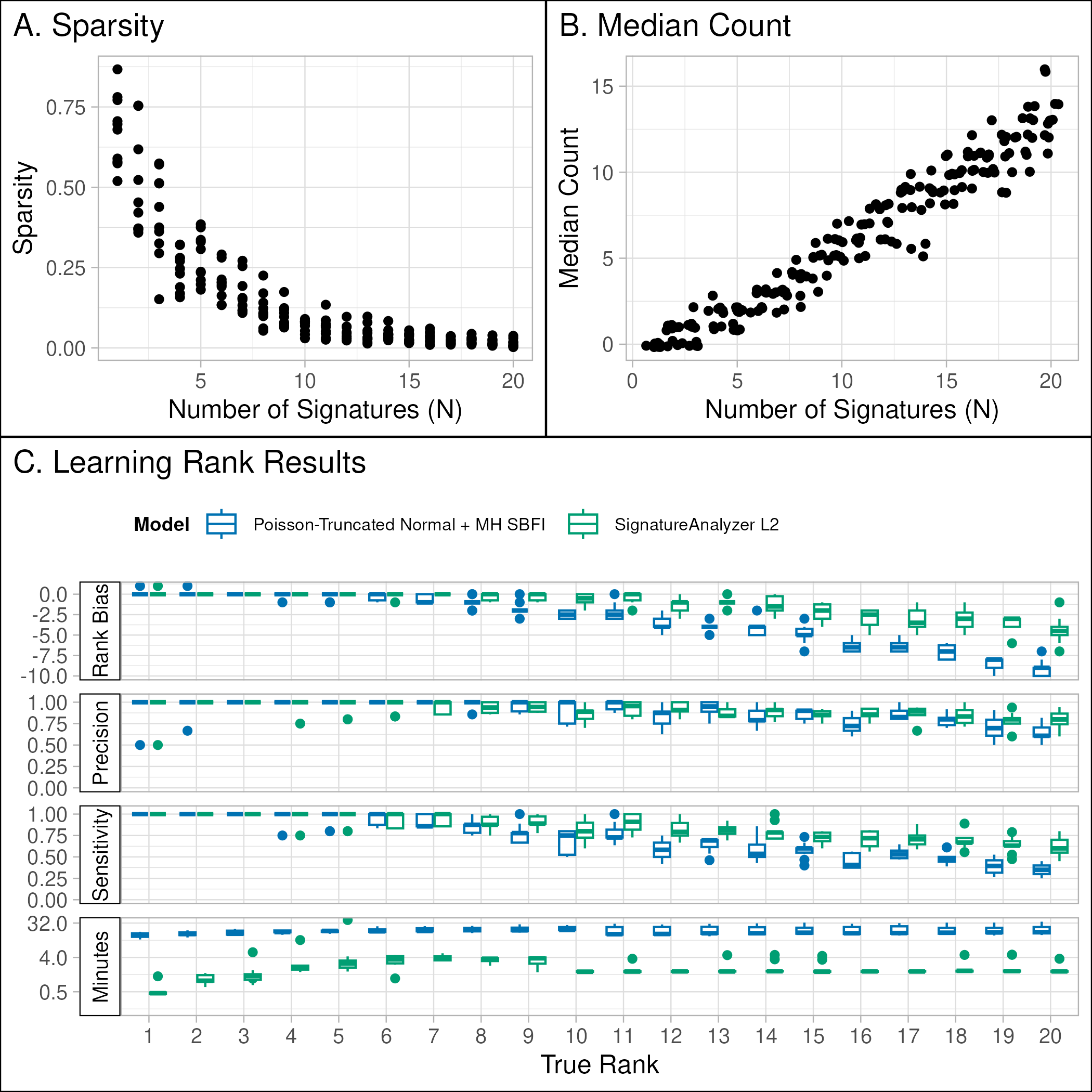}
    \caption{Data composition and results for learning rank with sparse data. \textbf{A.} Sparsity (proportion 0s) of simulated data. \textbf{B.} Median of mutational counts matrix of simulated datasets. \textbf{C.} Rank bias, precision, sensitivity, and time requirement (log scale) bayesNMF with Truncated Normal priors and SBFI compared to SignatureAnalyzer with L2 priors. Sensitivity is the proportion of true signatures for which there is an estimated signature with cosine similarity $>$0.9. Precision is the proportion of estimated signatures for which there is a true signature with cosine similarity $>$0.9.}
    \label{fig:study2_sparse}
\end{figure}

Data are simulated as in Section \ref{sec:sparse_fixed} with an expected 100 mutations per sample per signature. With N ranging from 1 to 20, sparsity ranges from 3-87\% sparse, where sparsity decreases with rank (Figure \ref{fig:study2_sparse}A). In a similar manner, the median mutational count increases with rank, ranging from 0-15 mutations (Figure \ref{fig:study2_sparse}B). This was done to maintain a relatively constant power to discover each signature (as was done in the primary analysis with expected 1000 per sample per signature).

Figure \ref{fig:study2_sparse}C compares rank accuracy, precision, and sensitivity between bayesNMF and SignatureAnalyzer. Our Poisson-Truncated Normal + MH with SBFI performs comparably to SignatureAnalyzer with L2 priors. However, for large ranks, SBFI might induce too much sparsity causing precision and sensitivity to drop below SignatureAnalyzer.

This sensitivity analysis shows that even with sparse data, bayesNMF is able to learn ranks between 1 and 10 with comparable precision and sensitivity to SignatureAnalyzer, though it is prone to underestimating high ranks more than SignatureAnalyzer.

\subsection{Sensitivity Analysis: Range of Learned Rank}

In the main text simulations learning rank, we specify the range 1-20 to bayesNMF. In this sensitivity analysis, we inspect performance when this range is expanded to 1-40 or narrowed to 1-10. We show that expanding the range does not bias learned rank upwards, and similarly that narrowing the range simply truncates estimated rank but does not bias estimated rank downward when the true rank is in the provided range.

\begin{figure}
    \centering
    \includegraphics[width=\linewidth]{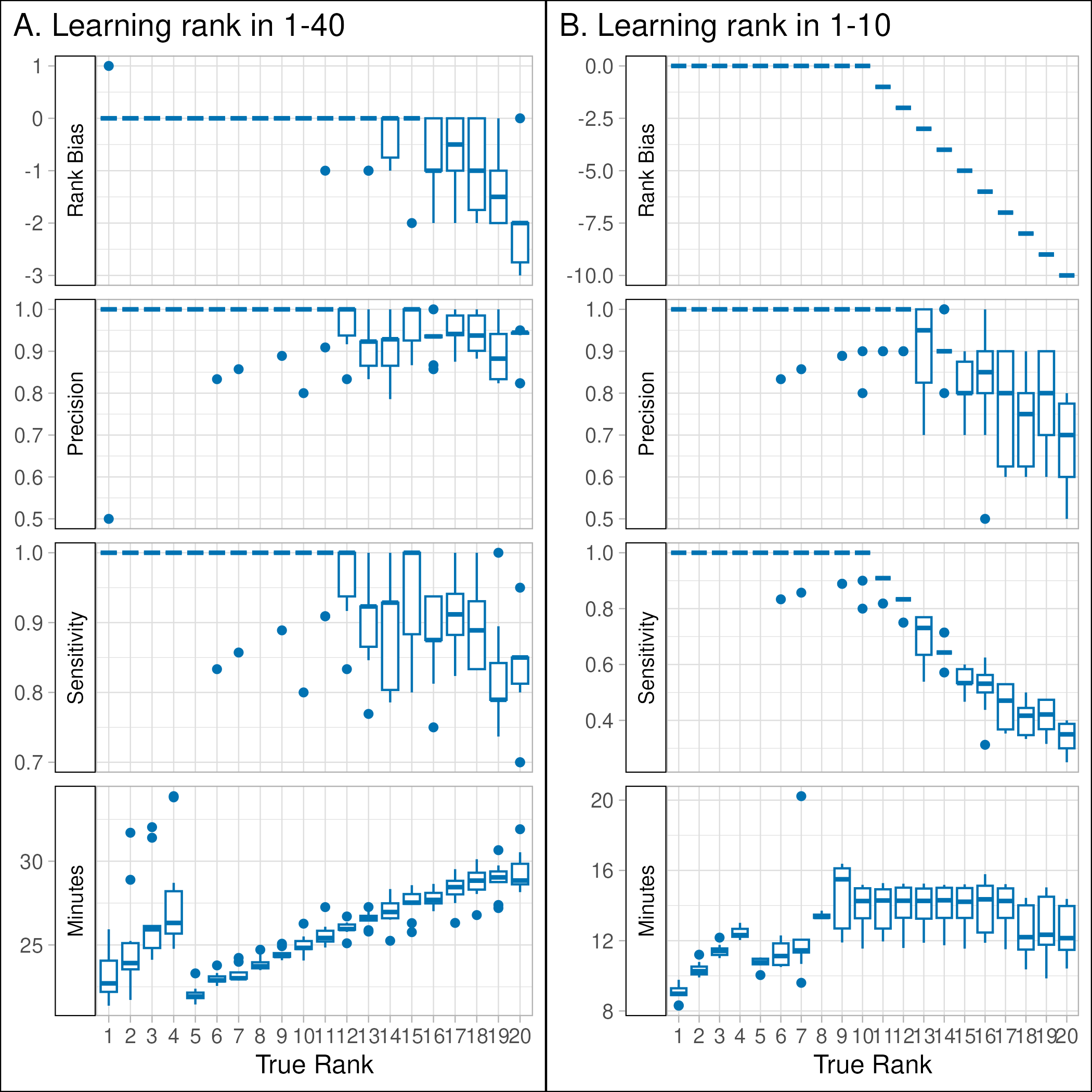}
    \caption{Rank bias, precision, sensitivity, and time for sensitivity analyses adjusting rank range. \textbf{A.} Results for {\tt rank = 1:40}. \textbf{B.} Results for {\tt rank = 1:10}.}
    \label{fig:study2_N10_N40}
\end{figure}

When the range is expanded to consider ranks 1-40, we see the same general results (with a bit of noise) in terms of rank bias, precision, and sensitivity (Figure \ref{fig:study2_N10_N40}A). This sensitivity analysis shows that increasing the maximum rank does not bias the learned rank upwards. It also shows that our model's underestimation of ranks 16-20 in the primary analysis is due to a lack of signal in the data, not a downward bias from specifying a maximum rank of 20.

When the range is narrowed to consider ranks 1-10,  we see the same general results (with a bit of noise) for true ranks 1-10 in terms of rank bias, precision, and sensitivity (Figure \ref{fig:study2_N10_N40}B). This means that decreasing the maximum rank does not bias the learned rank downwards, as long as the true rank is within the specified range. For ranks over 10, bayesNMF estimates the specified maximum rank of 10. This shows that when there is enough signal in the data, bayesNMF is able to learn that maximum rank if appropriate, reassuring us that our model's underestimation of ranks 16-20 is due to a lack of signal in the data, not a downward bias from the specified maximum rank. Further, this leads us to the conclusion that if the maximum rank is estimated, it is recommended to rerun the sampler with a higher maximum.

\newpage
\newpage

\section{Appendix D: Additional Data Application Results}
\label{supp_sec:appendix_d}

\subsection{Identifying Hypermutated Samples} \label{sec:hypermut}

\begin{figure}[b]
    \centering
    \includegraphics[width=0.8\linewidth]{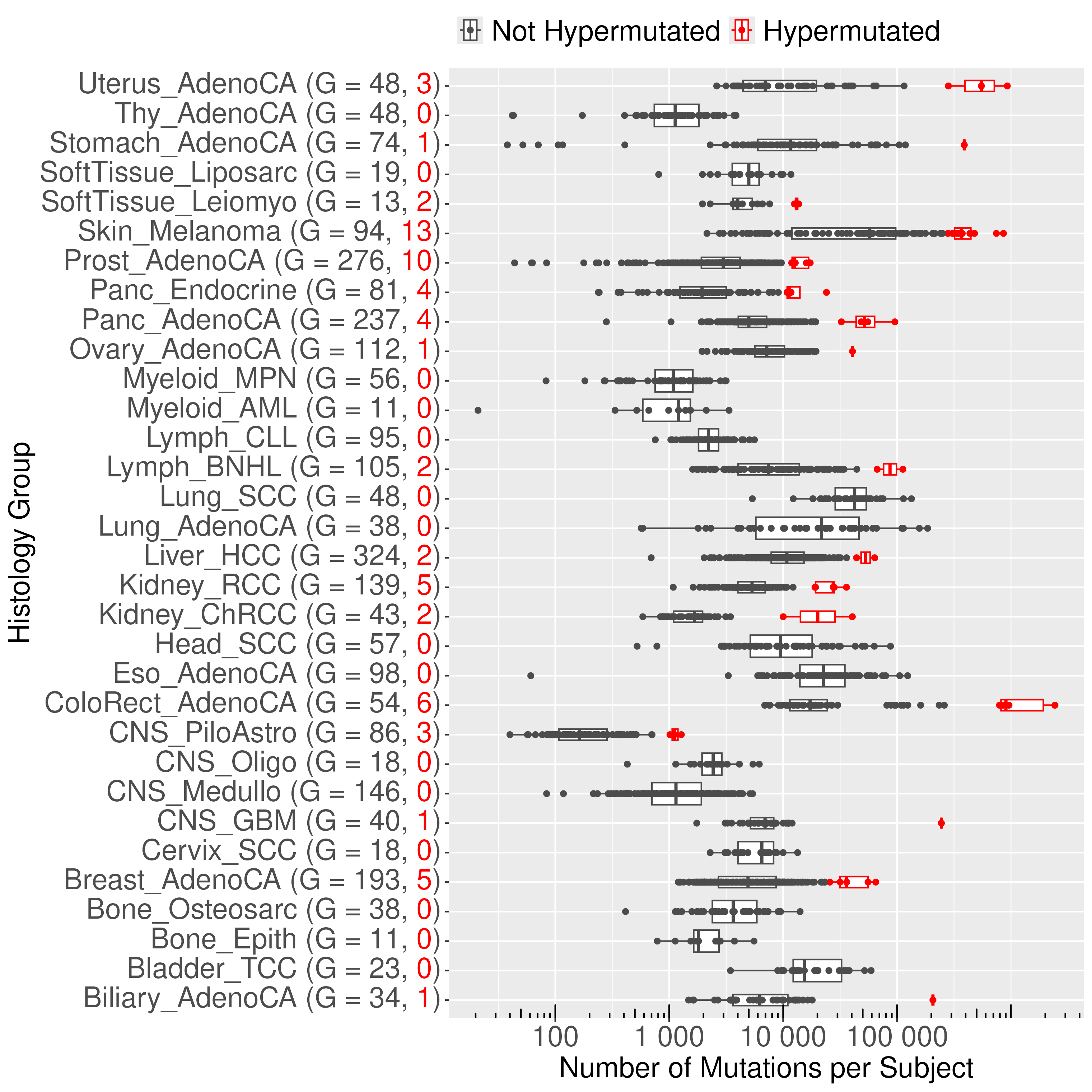}
    \caption{PCAWG data: total mutations per sample $m_g$ (log10 scale) plotted for each histology group and colored by hypermutation status. The y-axis labels report the number of cancer genomes $G$ separately for non-hypermutated (black) and hypermutated (red).}
    \label{fig:mut_counts_hypermutated_with_n}
\end{figure}

The total number of mutations per sample, $m_g = \sum_k M_{kg}$, is modeled by a negative binomial mixture model, which is fit using the Expectation-Maximization algorithm \citep{dempster1977maximum}. The number of mixture components, $C \in \{1, 2, ..., 10\}$ is chosen in a two-step fashion: first, if BIC is optimized at 1 component, we say there are no hypermutated samples, and otherwise, we choose the number of components between 2 and 10 to optimize the silhouette score. Samples are assigned to clusters based on the mixture component with the highest a-posteriori probability. Figure \ref{fig:mut_counts_hypermutated_with_n} shows the distribution of $m_g$ for each histology group as well as the final clustering into hypermutated and non-hypermutated samples.

\subsection{Additional Results}

\begin{figure}
    \centering
    \includegraphics[width = 0.9\linewidth]{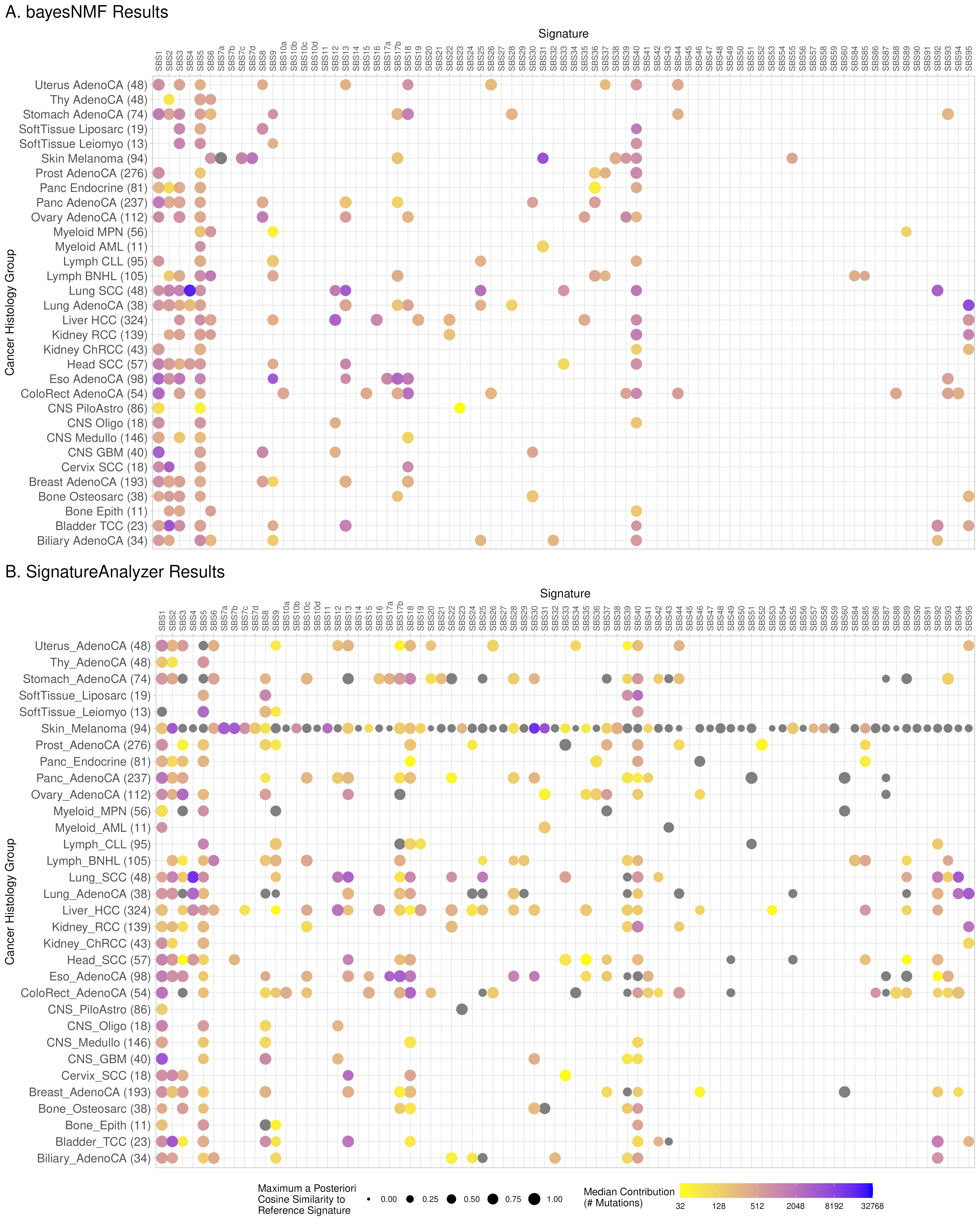}
    \caption{Results of {\tt bayesNMF}+SBFI and {\tt SignatureAnalyzer}+ARD on PCAWG SBS mutations data. Signatures identified in each histology group, sized by cosine similarity to the COSMIC reference signature and colored by median contribution among samples with nonzero mutations attributed to the signature (grey if median $<$1).  \textbf{A.} {\tt bayesNMF}+SBFI. \textbf{B.} {\tt SignatureAnalyzer}+ARD. (Figures inspired by \citet{alexandrov2020repertoire} Figure 3.)}
    \label{fig:suppD_pcawg_res}
\end{figure}

\begin{figure}
    \centering
    \includegraphics[width = \linewidth]{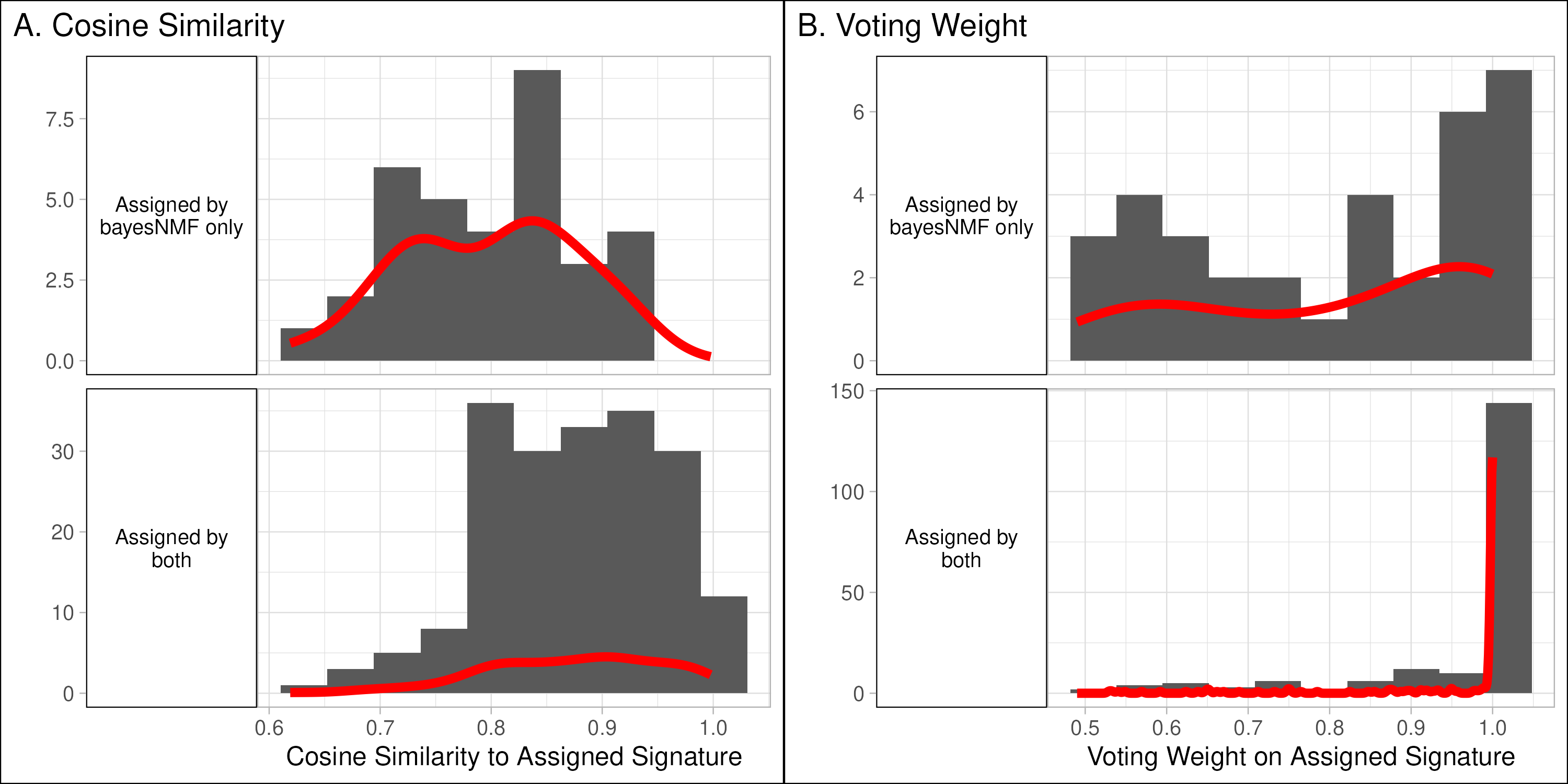}
    \caption{Uncertainty results on PCAWG data, separating signatures identified by both bayesNMF and SignatureAnalyzer and those identified by bayesNMF alone. \textbf{A.} Cosine similarity to assigned reference signature. \textbf{B.} Voting weight to assigned reference signature.}
    \label{fig:overlap}
\end{figure}

\begin{itemize}
    \item Figure \ref{fig:suppD_pcawg_res}: For PCAWG results using {\tt bayesNMF}+SBFI and {\tt SignatureAnalyzer}+ARD, this figure reports median number of mutations attributed to each signature and the cosine similarity of each estimated signature to its assigned reference signature (posterior average for bayesNMF).
    \item Table \ref{tab:ranks}: Reports estimated rank for PCAWG results using {\tt bayesNMF}+SBFI and {\tt SignatureAnalyzer}+ARD.
    \item Figure \ref{fig:overlap}: Investigates signatures assigned by both {\tt bayesNMF} and {\tt SignatureAnalyzer} versus signatures assigned by {\tt bayesNMF} only. The signatures assigned by {\tt bayesNMF} only have lower cosine similarity to the assigned reference and a lower proportion of assignment votes, indicating uncertainty in assignment.
\end{itemize}

\begin{table}[]
    \centering
    \scriptsize
    \begin{tabular}{|l|r|r|}
    \hline
    \textbf{Name} & \textbf{SignatureAnalyzer Rank}& \textbf{bayesNMF Rank}\\
    \hline
    Biliary AdenoCA & 12 & 9\\
    \hline
    Bladder TCC & 13 & 9\\
    \hline
    Bone Epith & 5 & 4\\
    \hline
    Bone Osteosarc & 9 & 7\\
    \hline
    Breast AdenoCA & 15 & 8\\
    \hline
    CNS GBM & 7 & 6\\
    \hline
    CNS Medullo & 5 & 4\\
    \hline
    CNS Oligo & 4 & 4\\
    \hline
    CNS PiloAstro & 2 & 3\\
    \hline
    Cervix SCC & 6 & 4\\
    \hline
    ColoRect AdenoCA & 27 & 14\\
    \hline
    Eso AdenoCA & 22 & 10\\
    \hline
    Head SCC & 18 & 9\\
    \hline
    Kidney ChRCC & 4 & 4\\
    \hline
    Kidney RCC & 10 & 7\\
    \hline
    Liver HCC & 27 & 11\\
    \hline
    Lung AdenoCA & 22 & 11\\
    \hline
    Lung SCC & 20 & 11\\
    \hline
    Lymph BNHL & 17 & 10\\
    \hline
    Lymph CLL & 7 & 5\\
    \hline
    Myeloid AML & 3 & 2\\
    \hline
    Myeloid MPN & 6 & 4\\
    \hline
    Ovary AdenoCA & 14 & 9\\
    \hline
    Panc AdenoCA & 18 & 9\\
    \hline
    Panc Endocrine & 9 & 6\\
    \hline
    Prost AdenoCA & 13 & 5\\
    \hline
    Skin Melanoma & 76 & 10\\
    \hline
    SoftTissue Leiomyo & 5 & 4\\
    \hline
    SoftTissue Liposarc & 4 & 4\\
    \hline
    Stomach AdenoCA & 27 & 11\\
    \hline
    Thy AdenoCA & 3 & 3\\
    \hline
    Uterus AdenoCA & 17 & 10\\
    \hline
    \end{tabular}
    \caption{Estimated ranks of {\tt bayesNMF}+SBFI and {\tt SignatureAnalyzer}+ARD on PCAWG SBS mutations data.}
    \label{tab:ranks}
\end{table}

\end{appendix}
\end{document}